\documentclass[fleqn,usenatbib]{mnras}

\usepackage{newtxtext,newtxmath}

\usepackage[T1]{fontenc}

\DeclareRobustCommand{\VAN}[3]{#2}
\let\VANthebibliography\thebibliography
\def\thebibliography{\DeclareRobustCommand{\VAN}[3]{##3}\VANthebibliography}

\usepackage{graphicx}	
\usepackage{amsmath}	
\usepackage[colorinlistoftodos]{todonotes}

\title[Cosmic-neighborhood distances to VHE BL Lacs]{Cosmic-neighbor-associated distances to VHE and candidate VHE BL Lacs via SDSS photometric redshifts}

\author[K. I. I. Koljonen et al.]{
Karri I. I. Koljonen,$^{1}$\thanks{E-mail: karri.koljonen@ntnu.no}
Elina Lindfors,$^{2, 3}$\thanks{E-mail: elina.lindfors@utu.fi}
\\
$^{1}$Institutt for Fysikk, Norwegian University of Science and Technology, H{\o}gskloreringen 5, Trondheim, 7491, Norway\\
$^{2}$ Department of Physics and Astronomy, Vesilinnantie 5, University of Turku, Finland\\
$^{3}$ Finnish Centre for Astronomy with ESO, Quantum, Vesilinnantie 5, University of Turku, Finland\\
}

\date{Accepted XXX. Received YYY; in original form ZZZ}

\pubyear{2015}

\begin{document}
\label{firstpage}
\pagerange{\pageref{firstpage}--\pageref{lastpage}}
\maketitle

\begin{abstract}
BL~Lac objects have distances that are difficult to estimate because their optical spectra rarely show identifiable emission or absorption lines. We present a method to constrain BL~Lac distances by associating them with their cosmic neighborhoods (local overdensities in the cosmic web) identified in optical fields. Distances to galaxies in these neighborhoods are estimated using the Sloan Digital Sky Survey (SDSS) photometric redshifts. Using cosmic-neighborhood associations previously established from multi-object spectroscopy and BL~Lacs with spectroscopically measured redshifts as control samples, we find that the redshifts of the associated neighborhoods can be recovered with an accuracy of $\Delta z \sim 0.02-0.04$. When multiple candidate neighborhoods are present in the field, we rank the candidate associations using a central-location criterion. In 50\% of cases, the BL~Lac is associated with the correct neighborhood. We provide distance estimates for BL~Lac objects without secure redshift measurements, focusing on sources detected at very high energies (VHE; $E>100$,GeV) and on candidate VHE emitters. Combining intervening absorption systems, host-galaxy imaging, $\gamma$-ray constraints, and our SDSS group-association method, we propose likely redshifts for TXS~0141$+$268 ($z\sim0.93-0.95$), PKS~0735$+$178 ($z=0.64$), 3FHL~J0905.5$+$1357 ($z=0.644$), B2~0912$+$29 ($z = 0.53$), 3FHL~J1120.8$+$4212 ($z=0.363$), BZB~J1243$+$3627 ($z=0.485$), 3FHL~J1253.1$+$5300 ($z=0.664$), and RX~J2156.0$+$1818 ($z=0.635$). Finally, we discuss how forthcoming narrow-band photometric-redshift surveys can improve blazar distance estimates via group-association techniques.
\end{abstract}

\begin{keywords}
galaxies: active -- BL Lacertae objects: general -- galaxies: distances and redshifts -- large-scale structure of Universe -- galaxies: groups: general -- methods: statistical
\end{keywords}



\section{Introduction}

Blazars are active galactic nuclei (AGN) with relativistic jets oriented very close to our line of sight. They are extremely bright across the electromagnetic spectrum and constitute the most numerous class of sources in the extragalactic $\gamma$-ray sky. The {\it Fermi} Large Area Telescope (LAT) has detected 3407 blazars above $50$\,MeV over a 12-year period \citep{2022ApJS..263...24A}. The number of known extragalactic very-high-energy (VHE) $\gamma$-ray sources, with photon energies above 100\,GeV, has increased from 10 to 85 over the past 15 years\footnote{\url{www.tevcat.org}, status November 2025}, and this trend is expected to accelerate with the advent of the Cherenkov Telescope Array Observatory (CTAO). VHE blazars are particularly valuable because they probe the most energetic particle-acceleration processes in the Universe; accurate distance estimates are crucial for determining intrinsic luminosities and for interpreting their emission mechanisms.

Blazars are commonly divided into two subclasses: flat-spectrum radio quasars (FSRQs) and BL~Lacertae objects (BL~Lacs). This division is traditionally based on their optical spectra. FSRQs show broad emission lines, while BL~Lacs are dominated by a nearly featureless synchrotron continuum, making spectroscopic redshift determination challenging. Redshift measurements generally rely on the observed emission lines thought to form in the narrow- or broad-line regions around the AGN and/or on absorption features from the host galaxy. In BL~Lacs, however, emission lines are intrinsically weak and the host-galaxy absorption features can be overwhelmed by the non-thermal continuum, falling below the noise level. Because redshift is required to derive intrinsic properties such as luminosity, this is a major obstacle for population studies of BL~Lacs.

Numerous observing programmes have targeted BL~Lacs \citep[e.g.][]{2006A&A...457...35S, shaw13, 2017ApJ...844..120P, 2020MNRAS.497...94P, 2021A&A...650A.106G, 2023MNRAS.518.2675K, dammando24, 2025A&A...704A.190R} using 10-meter-class telescopes to obtain high signal-to-noise spectra, yet many targets remain featureless, motivating alternative distance constraints. In the optical band, alternatives include host-galaxy imaging \citep[treating hosts as standard candles; e.g.][]{2003A&A...400...95N, 2005ApJ...635..173S, nilsson24} and intervening absorption systems \citep[e.g.][]{2013ApJ...768L..31F, 2022MNRAS.509.4330D}. $\gamma$-ray data can also provide constraints: \citet{2010MNRAS.405L..76P} used combined {\it Fermi}-LAT and VHE observations under the assumption that the intrinsic (extra-galactic-background-light-corrected) VHE spectrum cannot be harder than the LAT spectrum; related approaches have been applied widely \citep[e.g.][]{2023A&A...670A..49M, 2025A&A...694A.308M}. In addition, \citet{2021ApJ...920..118D, narendra22} used machine-learning methods based on {\it Fermi}-LAT properties and optical fluxes to estimate redshifts.

Here, we estimate BL~Lac distances from their cosmic neighbours. Group-based distance estimates have previously been derived from multi-object spectroscopic (MOS) observations \citep{rovero16, 2018MNRAS.474.3162T, rosagonzalez19, 2023A&A...680A..52P}. In our recent work \citep{2024MNRAS.531.5084K}, we showed that using photometric redshifts for galaxies surrounding the blazar yields results consistent with MOS. Building on that approach, we focus here on BL~Lacs detected at VHE and on candidate VHE emitters with uncertain or unknown redshifts, and we present new distance constraints for several $\gamma$-ray and VHE $\gamma$-ray-emitting blazars.

\section{Data and Sample} \label{sec:sample}

Several blazar catalogues contain thousands of BL~Lac objects, many without measured redshifts. For example, among the 3511 AGN listed in the 4LAC catalogue \citep{2020ApJ...892..105A}, only 1767 have measured redshifts (50.3\%). In this work we use data from Sloan Digital Sky Survey (SDSS) Data Release 19; therefore, all BL~Lacs in both our main sample and the control samples (Section~\ref{sec:control}) must lie within the SDSS footprint, limiting the accessible subset.

Our main sample is composed primarily of VHE BL~Lacs, supplemented by a smaller number of candidate VHE emitters selected from hard-spectrum {\it Fermi} catalogues and CTAO detectability studies. We compiled BL~Lacs with uncertain or unknown redshifts from several sources, focusing on objects for which improved distance constraints are particularly relevant for VHE studies. A first subset was taken from TeVCat\footnote{\url{https://www.tevcat.org}} including three VHE-detected BL~Lacs in the SDSS footprint with uncertain or poorly constrained redshifts: GB6~J1058$+$2817, 3FHL~J1120.8$+$4212, and PKS~1413$+$135. We then added candidate VHE emitters from \citet{2021A&A...650A.106G}, who evaluated CTAO detectability for hard-spectrum 3FHL blazars. This subset includes TXS~0141$+$268, 3FHL~J0905.5$+$1357, 3FHL~J1120.8$+$4212, 3FHL~J1150.5$+$4154, BZB~J1243$+$3627, 3FHL~J1253.1$+$5300, 3FHL~J1447.9$+$3608, and RX~J2156.0$+$1818. Finally, we include a small number of additional BL~Lacs with only imaging-based redshift estimates \citep{2003A&A...400...95N,nilsson12}, such as recently-identified VHE BL~Lac B2~0912$+$29 \citep{benbow25} and neutrino-candidate source PKS~0735$+$178 \citep[e.g.,][and references therein]{sahakyan23}. All objects in the main sample have previously published redshift limits or tentative estimates. Many have been targeted by deep spectroscopy that yielded lower limits (e.g., from intervening absorption systems or from non-detection of expected emission lines), and in a few cases a single emission line has led to tentative redshift assignments (see Section~\ref{sec:results}).

We define two control samples. The first consists of 16 BL~Lacs with confirmed spectroscopic redshifts (Table~\ref{tab:bcontrol}). These are taken from the MOJAVE hard-spectrum \textit{Fermi} AGN sample\footnote{\url{https://www.cv.nrao.edu/MOJAVE/MOJAVEhardspec.html}} and from \citet[][their Table~4]{nilsson24}. The second comprises of three BL~Lacs whose environments were studied with MOS (Table~\ref{tab:control}).

From SDSS DR19 we retrieve all available photometric and spectroscopic galaxy redshifts within a 2$\arcmin$ radius around each blazar. This corresponds to transverse physical scales of approximately 1.0, 1.5, and 1.8~Mpc at $z=0.25$, 0.5, and 0.75, respectively. The number of redshifts per field is reported in Appendix~\ref{sec:sdss_spec} (Table~\ref{tab:sdss_redshift}). In most fields, tens of galaxies have photometric redshifts but only a few have spectroscopic redshifts; spectroscopy alone is therefore insufficient to identify groups robustly. In Section~\ref{sec:method}, we describe our photometric-redshift group-identification and distance-estimation method.

\section{Photometric-redshift group distance estimation method}
\label{sec:method}

To characterise the redshift distribution of galaxies near each target and to identify candidate structures along the line of sight, we use a probabilistic procedure that \textit{i)} incorporates individual photometric-redshift uncertainties, \textit{ii)} constructs a non-parametric estimate of the underlying redshift distribution, and \textit{iii)} decomposes the line-of-sight distribution into into candidate structures with an error-aware Gaussian mixture model (GMM).

We select galaxies within 2$\arcmin$ from the blazar (Section~\ref{sec:sample}). For each galaxy we use its photometric redshift $z_i$ and uncertainty $\sigma_{z,i}$ to define a Gaussian probability density function (PDF) $p_i(z) = \mathcal{N}(z_i,\sigma_{z,i})$. 
We then compute the summed redshift distribution $P(z) = \sum_i p_i(z)$ on a fixed redshift grid for visualisation purposes (diagnostic plots; see Fig.~\ref{fig:pks1424}). The identification of candidate structures, however, is performed using an error-convolved mixture model that directly accounts for the heteroscedastic uncertainties $\sigma_{z,i}$.

We model the latent redshift distribution along the line of sight as a mixture of $K_{\rm max}=10$ components with means $\mu_k$ and weights $w_k$ ($\sum_k w_k=1$).\footnote{We adopt a relatively generous $K_{\rm max}$ and subsequently discard components with negligible effective membership and merge unresolved components (as described later), so the final number of reported structures is smaller in all fields.} 
Each observed SDSS photometric redshift $z_i$ is treated as a noisy measurement of the (unknown) true redshift, with a reported uncertainty $\sigma_{z,i}$. We approximate the per-galaxy redshift likelihood by a Gaussian kernel, $p(z_i\mid z_i^\ast)=\mathcal{N}(z_i\mid z_i^\ast,\sigma_{z,i}^2)$.
Assuming that the intrinsic dispersion of a galaxy group is negligible compared to SDSS photometric redshift uncertainties, the likelihood of the observed photometric redshift $z_i$ given mixture component $k$ with mean $\mu_k$ becomes
\begin{equation}
p(z_i \mid k)=\mathcal{N}\!\left(z_i \mid \mu_k,\, \sigma_{z,i}^2\right),
\end{equation}
and the mixture likelihood is
\begin{equation}
p(z_i)=\sum_{k=1}^{K_{\max}} w_k\,\mathcal{N}\!\left(z_i \mid \mu_k,\, \sigma_{z,i}^2\right).
\end{equation}
We fit this model via expectation--maximisation (EM) algorithm (see Appendix~\ref{app:em}), yielding posterior membership probabilities $r_{ik}=p(k\mid z_i)$ of galaxy $i$ in component $k$. For each fitted component we define an effective membership,
$N_k=\sum_i r_{ik}$,
and retain only components with $N_k \ge N_{\min}=1.99$, corresponding to at least a statistically significant pair of galaxies. In addition, we merge components whose mean redshifts differ by less than a field-dependent resolvability threshold,
\begin{equation}
\Delta z_{\mathrm{res}} = f\times\mathrm{median}(\sigma_{z,i}),
\end{equation}
with $f=0.7$. Merged components are constructed by summing $r_{i,\mathrm{merged}}=\sum r_{ik}$ and re-computing component summary quantities. The resulting merged components define our candidate redshift structures for each field.

To associate candidate redshift structures with a sky position, we compute a weighted centroid for each component:
\begin{equation}
\alpha_k = \frac{\sum_i r_{ik}\,\alpha_i}{\sum_i r_{ik}},\qquad
\delta_k = \frac{\sum_i r_{ik}\,\delta_i}{\sum_i r_{ik}}.
\end{equation}
We then compute the angular separation between the blazar and each component centroid,
$\Delta\theta_k=\mathrm{angdist}((\alpha_k,\delta_k),(\alpha_{\mathrm{src}},\delta_{\mathrm{src}}))$.
This separation provides an additional ranking metric because BL~Lacs are expected to lie preferentially near the centres of their host groups \citep{2019ApJS..240...20M, 2020ApJ...900L..34M, 2020ApJS..247...71M}.

The method is accompanied by a diagnostic plot (shown for all fields in the figures cited throughout the paper; see example in Fig. \ref{fig:pks1424}) that visualises both the summed redshift PDF $P(z)$ and the fitted mixture decomposition. The plot shows the individual galaxy PDFs $p_i(z)$ (thin black lines), their sum $P(z)$ (thick black line), the retained (and merged) mixture components (dashed coloured lines), and the total mixture model (red line). 
The legend reports the number of galaxies, the number of retained components, and for each component its mean redshift, the angular separation between its composite centroid and the blazar, and the effective number of galaxies in that group. Literature redshift measurement or constraints are indicated by vertical dotted lines (limits are shown with arrows). The source name and coordinates are shown in the upper-left corner.

We note that our method differs significantly from those developed to build galaxy-cluster catalogues from SDSS data. For example, \cite{2012ApJS..199...34W} uses SDSS photometric redshifts to identify overdense regions via a grouping algorithm, with the cluster redshift computed as the median redshift of the assigned member galaxies. This algorithm requires more galaxies than our method to flag a group as a cluster; consequently, most of the groups we identify are not included in the SDSS cluster catalogue of \cite{2012ApJS..199...34W}.

\section{Performance of the method} \label{sec:control}

In this section, we test our method using blazars with redshifts known from spectroscopy (Section~\ref{sec:spec_control}) and fields with galaxy groups confirmed by MOS (Section~\ref{sec:mos_control}). We also provide an exploratory comparison with narrow-band photometric-redshift surveys (Section~\ref{sec:survey_control}) to illustrate the precision that can be achieved with forthcoming data sets.

\subsection{Redshifts from blazar spectroscopy} \label{sec:spec_control}

\begin{table*}
    \caption{Control blazars with spectroscopic redshifts within the SDSS footprint. Columns: (1) blazar identifier; (2) reference for the published spectroscopic redshift estimate; (3) literature-reported spectroscopic redshift estimate; (4) redshift corresponding to the centroid of the GMM component most likely associated with the blazar (from our analysis); (5) difference between the group redshift and the literature value;  (6) angular separation between the blazar and the composite central position of the corresponding GMM component; (7) the effective number of galaxies in the group; (8) rank order of the group composite centroid in the field (I = closest to the blazar).}
    \label{tab:bcontrol}
    \centering
    \begin{tabular}{l|cccccccccc}   
    \hline
    Source & Ref & Spectro-z & Group-z & $\Delta$z & $\Delta\theta$('') & $N_k$ & Group order  \\
    \hline
    RGB~J0013$+$191  & 1 & 0.48 & 0.51 & +0.03 & 6  & 15.4 & I \\
    RGB~J0115$+$253  & 2 & 0.38 & 0.42 & +0.04 & 8  & 32.9 & I \\  
    PKS~0139$-$09    & 3 & 0.73 & 0.63 & -0.10 & 14 & 10.7 & III \\
    GB6~J0154$+$0823 & 1 & 0.68 & 0.67 & -0.01 & 34 & 3.3  & III \\
    IVS~B0200$+$30A  & 1 & 0.76 & 0.73 & -0.03 & 7  & 3.7  & I \\
    RGB~J0202$+$088  & 4 & 0.63 & 0.62 & -0.01 & 7  & 15.3 & II \\
    RGB~J0227$+$020  & 5 & 0.46 & 0.41 & -0.05 & 16 & 63.3 & I \\     
    RGB~J0757$+$099  & 2 & 0.27 & 0.30 & +0.03 & 26 & 11.7 & V \\
    1ES~0806$+$524   & 1 & 0.14 & 0.24 & -0.10 & 39 & 6.4  & III \\
    PKS~0823$+$033   & 1 & 0.50 & 0.50 & -0.00 & 7  & 8.3  & I \\
    1ES~1011$+$496   & 6 & 0.21 & 0.21 & +0.00 & 6  & 18.4 & I \\
    ON~325           & 7 & 0.13 & 0.20 & +0.07 & 24 & 8.3  & III \\
    W~Comae          & 7 & 0.10 & 0.11 & +0.01 & 53 & 6.2  & IV \\
    RGB~J1415$+$485  & 8 & 0.50 & 0.44 & -0.06 & 8  & 25.1 & I \\
    OX~183           & 1 & 0.88 & 0.95 & +0.07 & 18 & 2.5  & IV \\
    CTD~135          & 4 & 0.79 & 0.79 & -0.00 & 28 & 8.4  & IV \\
    \hline                
    \end{tabular} \\
    \textbf{References:} 1) \citet{shaw13}, 2) \citet{lamura22}, 3) \citet{rector01}, 4) \citet{shaw12}, 5) \citet{sbarufatti05}, 6) \citet{albert07}, 7) \citet{paiano17}, 8) SDSS DR13
\end{table*}

We first test the method on blazars with spectroscopic redshifts available in the literature. We recover the corresponding group redshifts with a typical accuracy of $\Delta z \sim 0.04$. The correct group centre is the closest one to the blazar in seven out of the 16 targets (Table~\ref{tab:bcontrol}; diagnostic plots in Appendix~\ref{sec:control_plots}, Figs.~\ref{fig:control_spec1} and \ref{fig:control_spec2}). In these cases, the angular separation between the blazar and the composite group centre ranges from 6$\arcsec$ to 16$\arcsec$.
In four additional cases (PKS~0139$-$09, RGB~J0202$+$088, RGB~J1415$+$485, and OX~183), the identified group centre lies within 18$\arcsec$ of the blazar, although it is not the nearest centroid in the field. 

For nearby blazars ($z<0.15$; 1ES~0806$+$524, ON~325 and W~Comae), the angular separation to the associated group centroid can appear comparatively large because our fixed $2\arcmin$ search radius probes physical distances smaller than 1~Mpc (e.g., at $z \approx 0.1$, $2\arcmin \approx 0.5$\,Mpc). Conversely, at high redshift ($z\gtrsim0.8$; OX~183 and CTD~135), the limited depth of SDSS imaging (complete to $r \approx 22$~mag) implies that only a small number of galaxies contribute to the high-$z$ GMM components. Because the galaxy catalogues are not complete in any given field, group identifications---especially at low and high redshift---should be interpreted with appropriate caution.

\subsection{Redshifts from MOS studies of blazar environments} \label{sec:mos_control}

We further test the method using MOS studies of blazar environments within the SDSS footprint: PKS~1424$+$240 \citep[$z = 0.601$;][]{rovero16}, PG~1553$+$113 \citep[$z = 0.433$;][]{johnson19}, and RGB~J2243$+$203 \citep[$z = 0.528$;][]{rosagonzalez19}. All spectroscopically-identified galaxy groups have counterparts among the components of our GMM analysis (see Fig.~\ref{fig:pks1424}, and Figs. \ref{fig:1es1553} and \ref{fig:rgs2243} in Appendix~\ref{sec:mos_plots}). The group redshifts are recovered with an accuracy of $\Delta z \sim 0.02$ (Table~\ref{tab:control}). The correct group centre is the closest centroid to the blazar for PKS~1424$+$240 and PG~1553$+$113, whereas for RGB~J2243$+$203 it corresponds to the third-closest centroid, but located just $\sim$13$\arcsec$ from the blazar position.

\begin{table*}
    \caption{Control fields with spectroscopically confirmed galaxy groups (from MOS studies). The group associations adopted in this work are indicated in bold. Columns: (1) blazar field; (2) literature reference for the MOS study; (3) spectroscopic redshift of the galaxy group; (4) number of galaxies identified as group members from spectroscopy; (5) redshift corresponding to the centroid of the GMM component most likely associated with the blazar (from our analysis); (6) difference between the group redshift and the literature value; (7) angular separation between the blazar and the composite central position of the corresponding GMM component; (8) the effective number of galaxies in the group; (9) rank order of the group composite centroid in the field (I = closest to the blazar).} 
    \label{tab:control}
    \centering
    \begin{tabular}{c|cccccccc}   
    \hline
    Field & Ref & $z$ & N & Group-z & $\Delta$z & $\Delta\theta$('') & $N_k$ & Group order \\
    \hline
     PKS~1424$+$240 & 1 & \textbf{0.60} & \textbf{8} & \textbf{0.61} & \textbf{+0.01} & \textbf{8} & \textbf{12.4} & \textbf{I} \\
                   &   & 0.47 & 2 & 0.49 & +0.02 & 10 & 31.7 & II \\
                   &   & 0.12 & 3 & 0.13 & +0.01 & 16 & 2.3 & III \\
                   &   & -- & -- & 0.21 & -- & 31 & 8.1 & IV \\
                   &   & -- & -- & 0.67 & -- & 34 & 3.6 & IV \\
                   &   & -- & -- & 0.84 & -- & 49 & 3.9 & V \\ 
    \hline
    1ES~1553$+$113 & 2 & 0.39/\textbf{0.43}$^{\dagger}$ & 8/\textbf{14} & \textbf{0.40} & +0.01/\textbf{-0.03} & \textbf{16} & \textbf{37.7} & \textbf{I} \\
                   &   & --        & --   & 0.73 & --        & 20 & 2.7 & II  \\
                   &   & 0.53/0.57 & 8/4  & 0.56 & +0.03/-0.01 & 24 & 11.9 & III \\
                   &   & --        & --   & 0.85 & --        & 29 & 8.1 & IV \\
                   &   & 0.15      & 5    & 0.11 & -0.04     & 49 & 2.6 & V \\
    \hline
    RGB~J2243$+$203 & 3 & --        & --   & 0.62  & --       & 7  & 10.5 & I \\
                    &   & --        & --   & 0.37  & --       & 12 & 21.3 & II \\
                    &   & \textbf{0.53}      & \textbf{4}    & \textbf{0.51} & \textbf{-0.02}      & \textbf{13} & \textbf{16.4} & \textbf{III} \\
                    &   & --        & --   & 0.28 & --        & 21 & 8.4 & IV \\     
                    &   & --        & --   & 0.78 & --        & 25 & 3.2 & V \\
                    &   & --        & --   & 0.19 & --        & 28 & 2.6 & VI \\
                    &   & --        & --   & 0.94 & --        & 36 & 3.1 & VII \\
                    &   & --        & --   & 0.08 & --        & 46 & 2.5 & VIII \\
    \hline                
    \end{tabular}\\
    $^{\dagger}$ Two values given on this row and the next correspond to two different galaxy groups found in \citet{johnson19}, with the blazar associated at $z=0.43$ (corresponds to the values in bold). \\
    \textbf{References:} 1) \citet{rovero16}, 2) \citet{johnson19}, 3) \citet{rosagonzalez19}    
\end{table*}

\begin{figure*}
 \centering
 \includegraphics[width=0.95\linewidth]{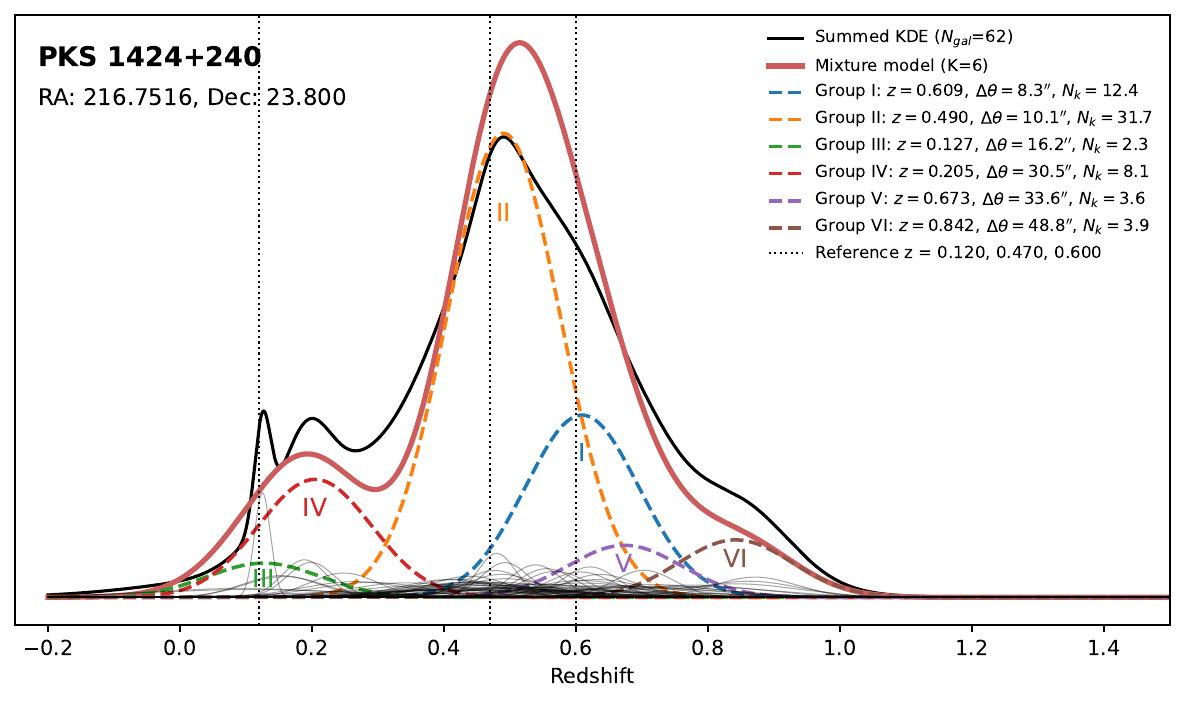}
 \caption{Redshift distribution of the SDSS galaxies in the field of PKS~1424$+$240. The summed KDE (black line), the total GMM model (red line), individual GMM components (dashed lines), and spectroscopic redshifts of the galaxy groups from the literature (dotted vertical lines). Each component is labeled with its mean redshift, width, and angular separation from the target source.   
 }
 \label{fig:pks1424}
\end{figure*} 

\subsection{Applicability to upcoming photometric surveys} \label{sec:survey_control}

Next, we compare our results with expectations from narrow-band photometric-redshift surveys, specifically the Physics of the Accelerating Universe Survey \citep[PAUS;][]{benitez09,padilla19} and the Javalambre Physics of the Accelerating Universe Astrophysical Survey \citep[J-PAS;][]{benitez14,bonoli21}. These surveys are designed to deliver near-spectroscopic photometric-redshift precision for large-scale structure studies through the use of multiple narrow-band filters. 
As discussed below, they enable substantially improved precision in identifying galaxy groups co-located with blazars than is possible with SDSS data. 

At present, public PAUS and J-PAS releases cover only a small fraction of the sky and do not include VHE blazars (or VHE candidates) with uncertain redshifts. We therefore  focus on blazars within the PAUS and J-PAS footprints that have known redshifts, in order to demonstrate the potential of narrow-band data sets in detecting associated galaxy groups around blazars with unknown redshifts.

\subsubsection{PAUS}

PAUS, carried out with the 4.2\,m William Herschel Telescope at the Roque de los Muchachos Observatory, employs 40 narrow-band filters spanning 450--850\,nm (FWHM~$\approx$~130\,\AA) across the 1~deg$^2$ PAUCam field of view, reaching $i \approx 23$. This configuration yields typical photometric-redshift precisions of $\Delta z_p = 0.003-0.02$ for $z < 1$ in deep extragalactic fields \citep{marti14}.

We use the PAUS Master Catalogue (PAUS-MC\footnote{\url{https://pausurvey.org/public-data-release/}}), which compiles narrow-band photometry for nearly two million detections over $\sim$50 deg$^2$ across several extragalactic fields (including COSMOS, W1 and W3 from CFHTLenS, and G09 from KIDS/GAMA). We cross-matched the CAZ blazar catalogue \citep{kouch25} to PAUS-MC and obtained six matches with $0.005<z<1$. For each blazar we considered a projected radius of 1~Mpc and constructed summed redshift distributions using a redshift-dependent widths for the PDFs consistent with the survey photo-$z$ precision. Specifically, we adopt $\Delta z_p=0.003$ for $z<0.33$, $\Delta z_p=0.008$ for $0.33\leq z<0.54$, $\Delta z_p=0.011$ for $0.54\leq z<0.67$, $\Delta z_p=0.019$ for $0.67\leq z<0.73$, $\Delta z_p=0.022$ for $0.73\leq z<0.92$, and $\Delta z_p=0.028$ for $z>0.92$. 

The resulting PAUS redshift distributions are shown in Appendix \ref{sec:paus_plots} (Figs~\ref{fig:paus_results_1} and \ref{fig:paus_results_2}). In all cases, the blazar redshift coincides with a galaxy overdensity, typically within $\Delta z \approx 0.001$. The corresponding weighted group-centroid offsets range from 10\arcsec to 63\arcsec, i.e. 62--151 kpc at their respective redshifts. 

We also analyse the same fields using SDSS data to and present the SDSS diagnostic plots alongside the PAUS distributions. The results are summarised in Table~\ref{tab:pau_redshift}. Using SDSS photometric redshifts, the group-based estimate agrees with the blazar redshift to within $\Delta z \lesssim 0.04$ on average, and the associated centroid is the closest one to the blazar in three out of six of cases.

\subsubsection{JPAS-EDR}

J-PAS, conducted with the 2.5\,m Javalambre Survey Telescope (JST/T250), uses 54 narrow-band filters (FWHM~$\approx$~145\,\AA) to cover $\sim$8500~deg$^2$ to a limiting magnitude of $r \approx 23.5$. Its design delivers photometric-redshift uncertainties of $\Delta z_p = 0.003(1+z)$ over $0.1 < z < 1.2$ \citep{laur22}.

We use data from the J-PAS Early Data Release (EDR), which covers $\sim$12~deg$^2$ in two regions observed with the full filter set and provides photometry for $\sim 6\times10^5$ sources. Given the limited sky coverage (and partial overlap with the G09 field), we find a single match between the CAZ blazar catalogue and the J-PAS EDR: BZB~J1631$+$4217. The corresponding results are included in Table~\ref{tab:pau_redshift} and shown in Fig.~\ref{fig:paus_results_2}.

\section{Cosmic-neighbor-associated redshifts to blazars} \label{sec:results}

\begin{figure*}
  \centering 
  \includegraphics[width=0.95\textwidth]{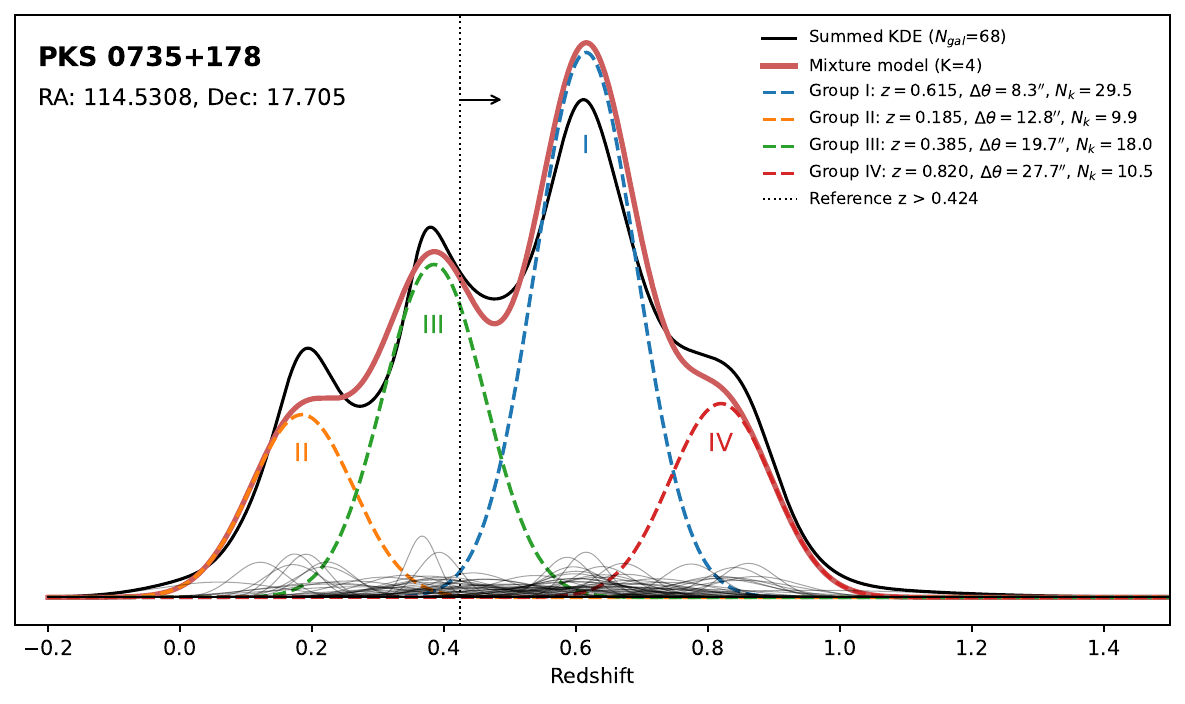}
  \caption{Redshift distribution of the SDSS galaxies in the field of PKS~0735$+$178. See Figure \ref{fig:pks1424} and Section \ref{sec:method} for explanation of the data. If the blazar is associated with a galaxy group, its most plausible redshift is $z\approx0.64$ (see discussion in Section \ref{sec:notes}).} \label{fig:blazar_res1}
\end{figure*}

\begin{figure*}
  \centering 
  \includegraphics[width=0.49\textwidth]{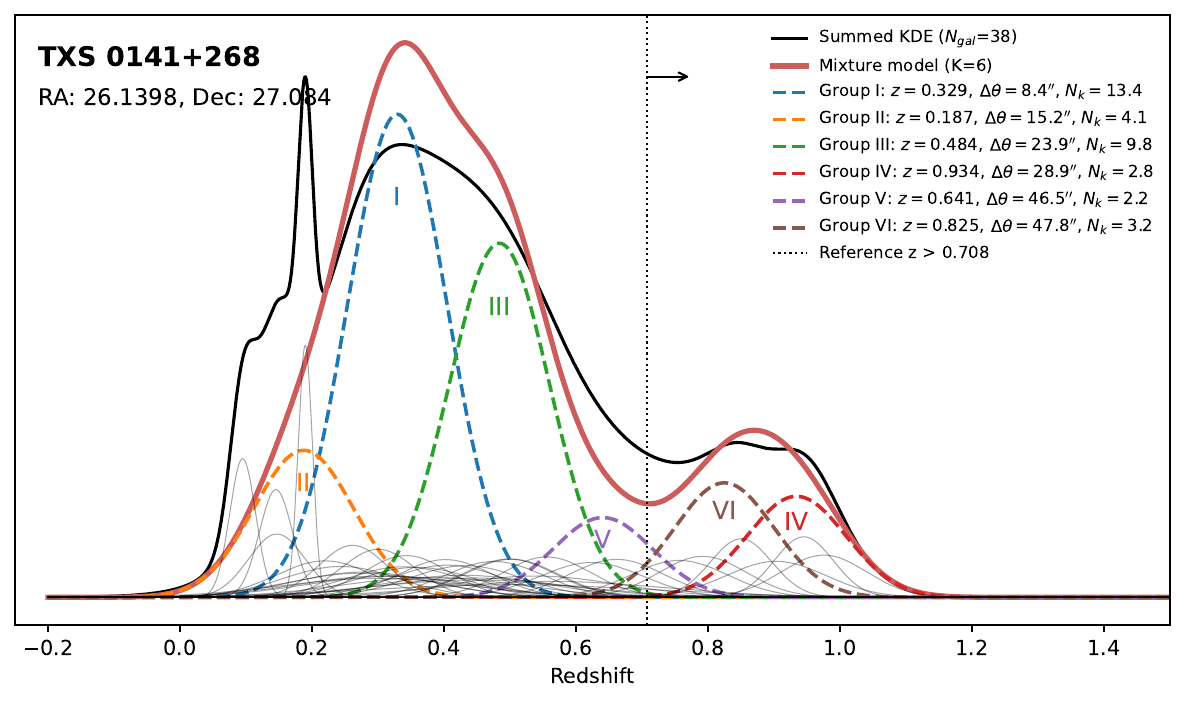}
  \includegraphics[width=0.49\textwidth]{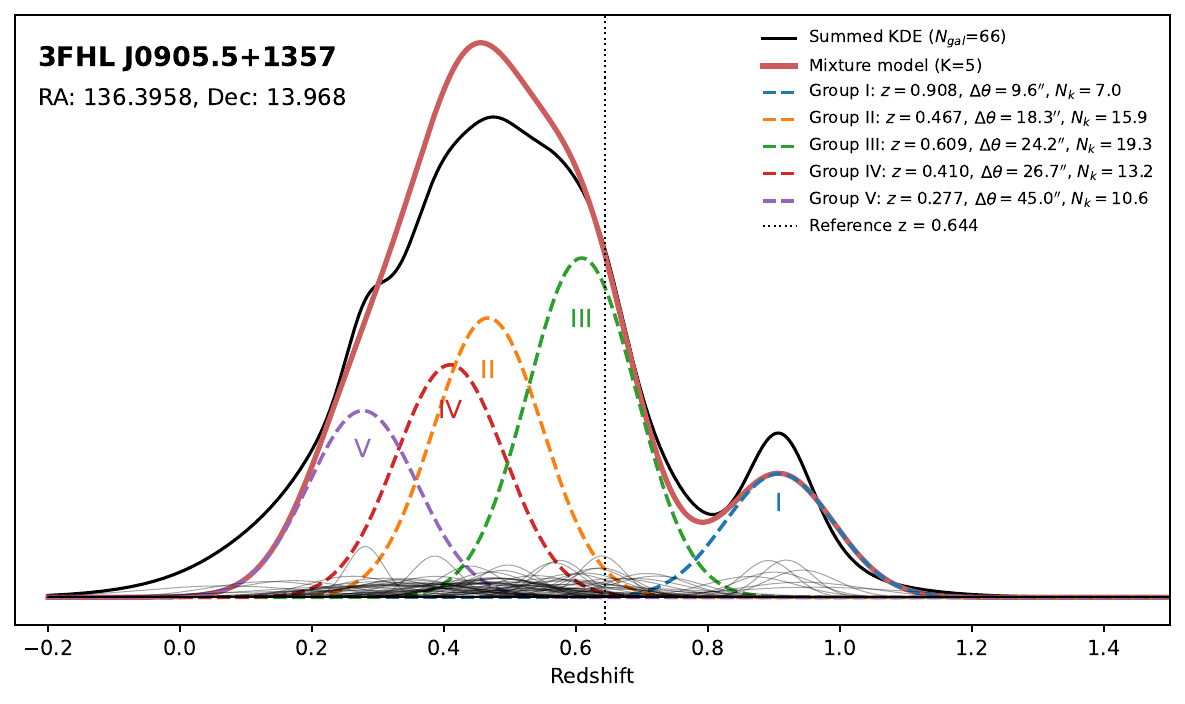}
  \includegraphics[width=0.49\textwidth]{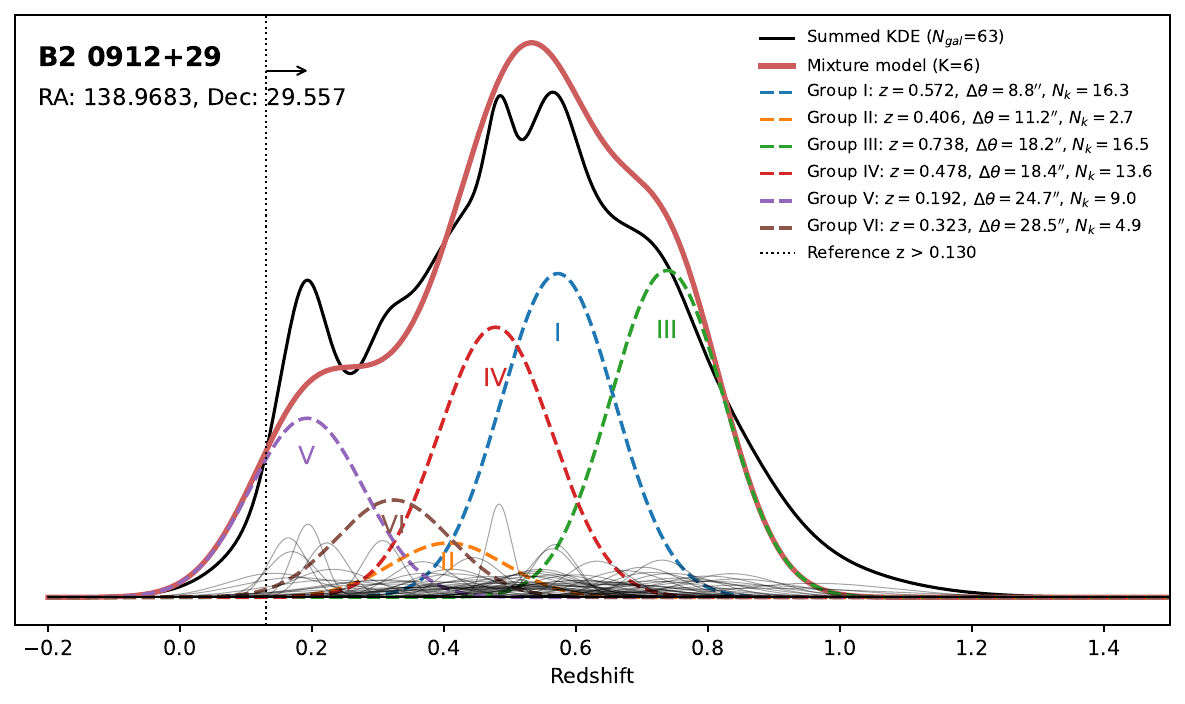}
  \includegraphics[width=0.49\textwidth]{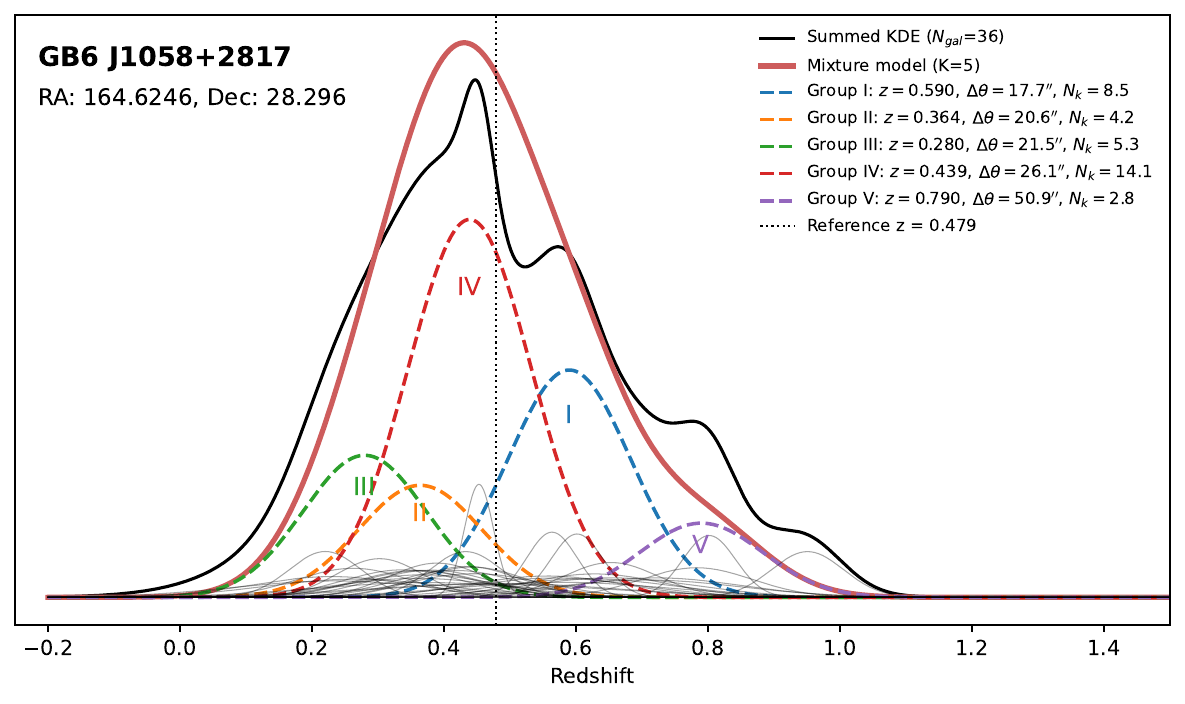}
  \includegraphics[width=0.49\textwidth]{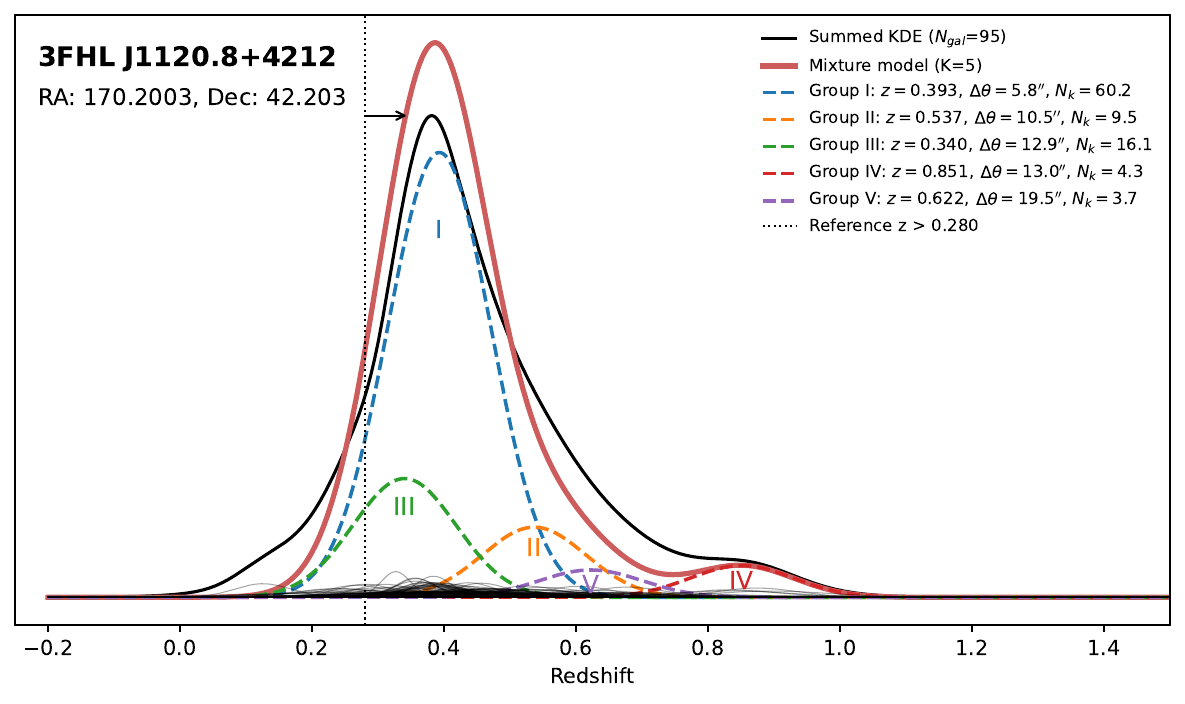}
  \includegraphics[width=0.49\textwidth]{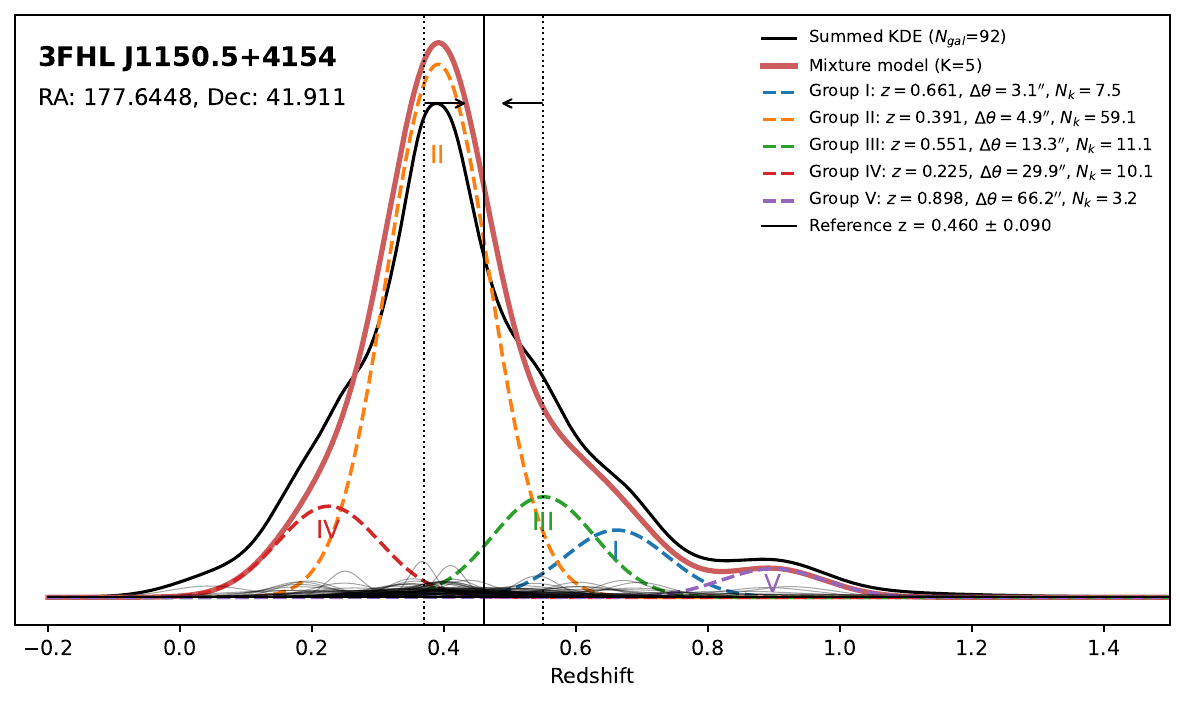}
  \includegraphics[width=0.49\textwidth]{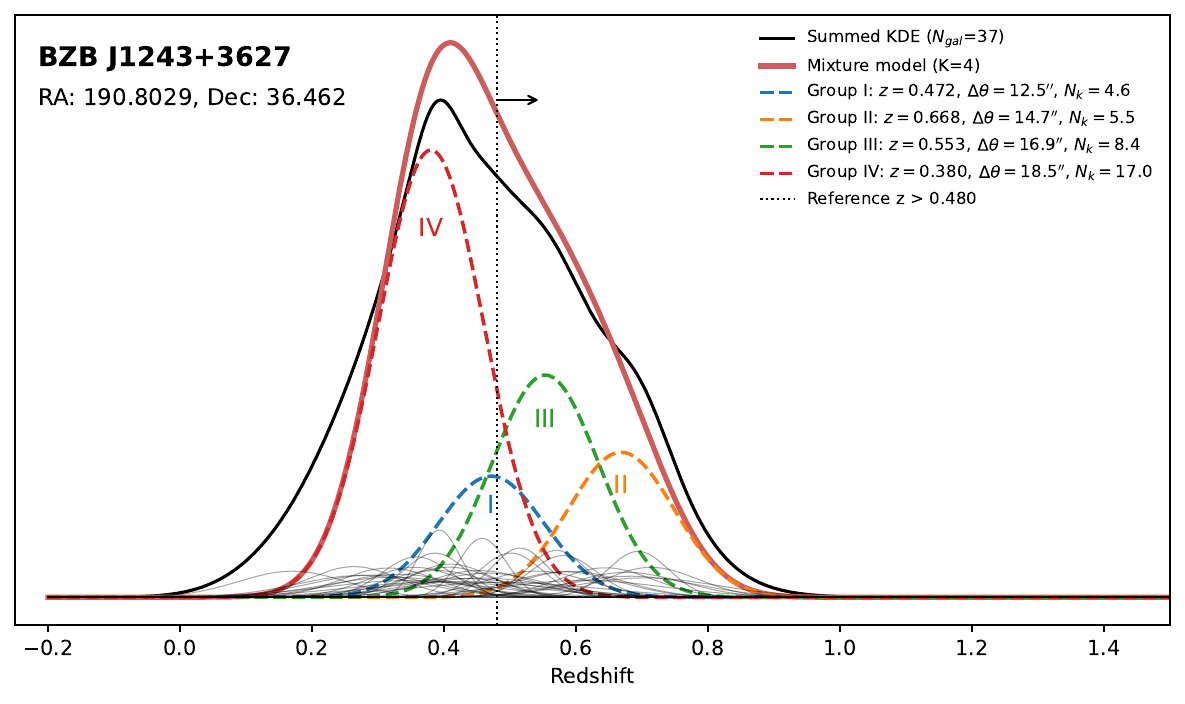}  
  \includegraphics[width=0.49\textwidth]{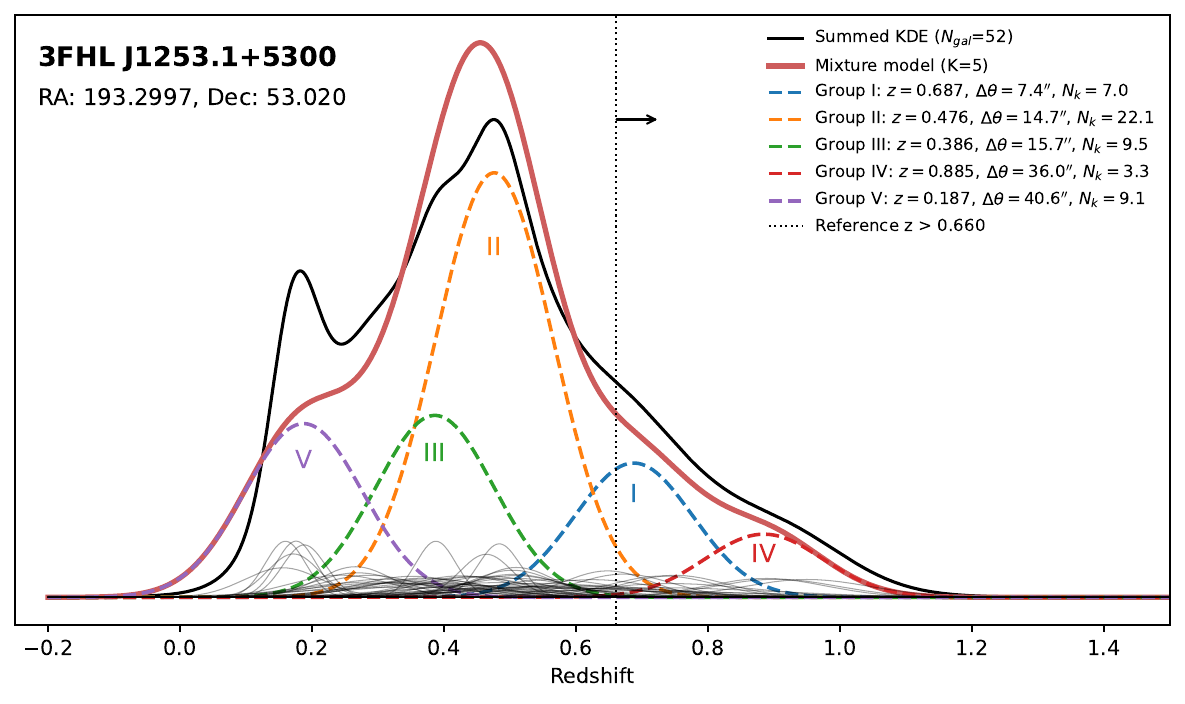}
  \caption{Redshift distribution of the SDSS galaxies in the fields of TXS~0141$+$268, 3FHL~J0905.5$+$1357, B2~0912$+$29, GB6~J1058$+$2817, 3FHL~J1120.8$+$4212, 3FHL~J1150.5$+$4154, BZB~J1243$+$3627, 3FHL~J1253.1$+$5300, and PKS~1413$+$135. See Figure \ref{fig:pks1424} and Section \ref{sec:method} for explanation of the data.} \label{fig:blazar_res2}
\end{figure*}

\begin{figure*}
  \centering
  \includegraphics[width=0.49\textwidth]{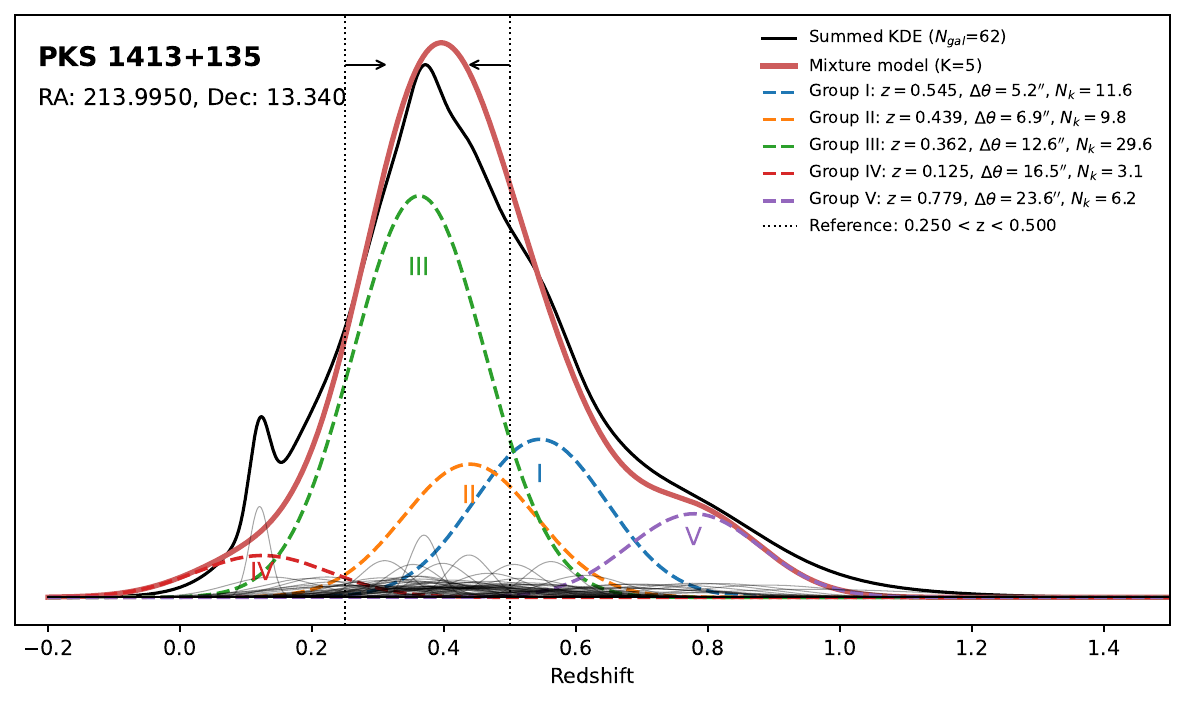}  
  \includegraphics[width=0.49\textwidth]{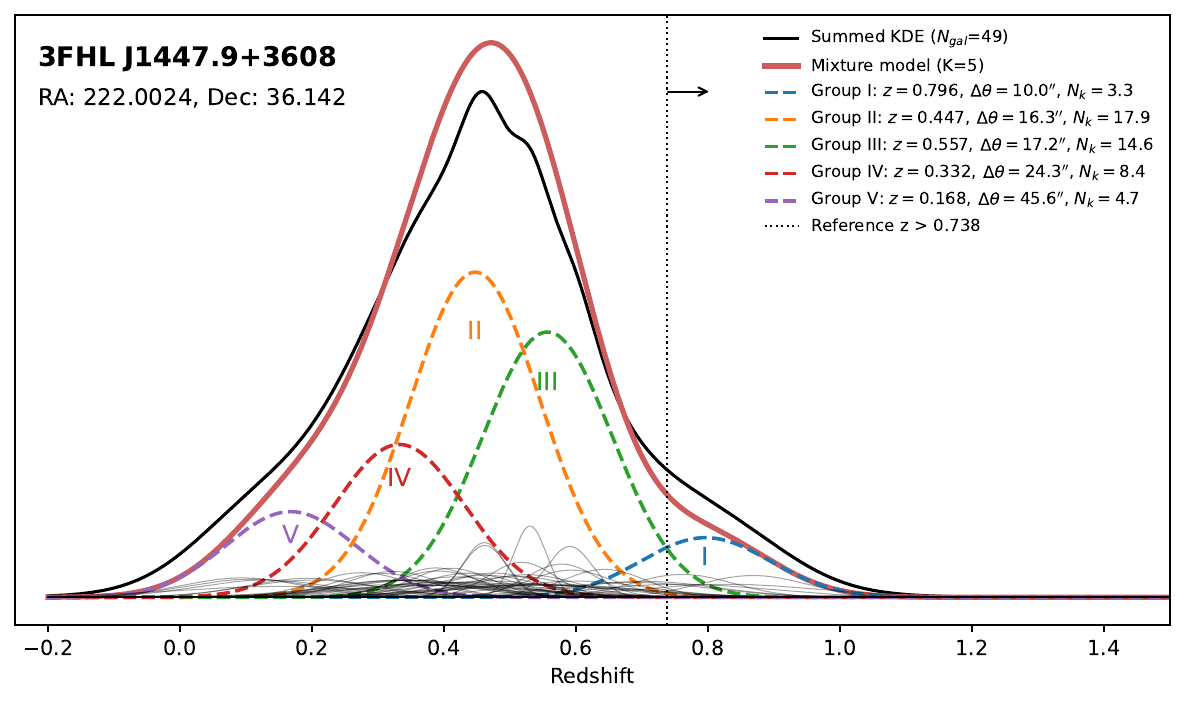}
  \includegraphics[width=0.49\textwidth]{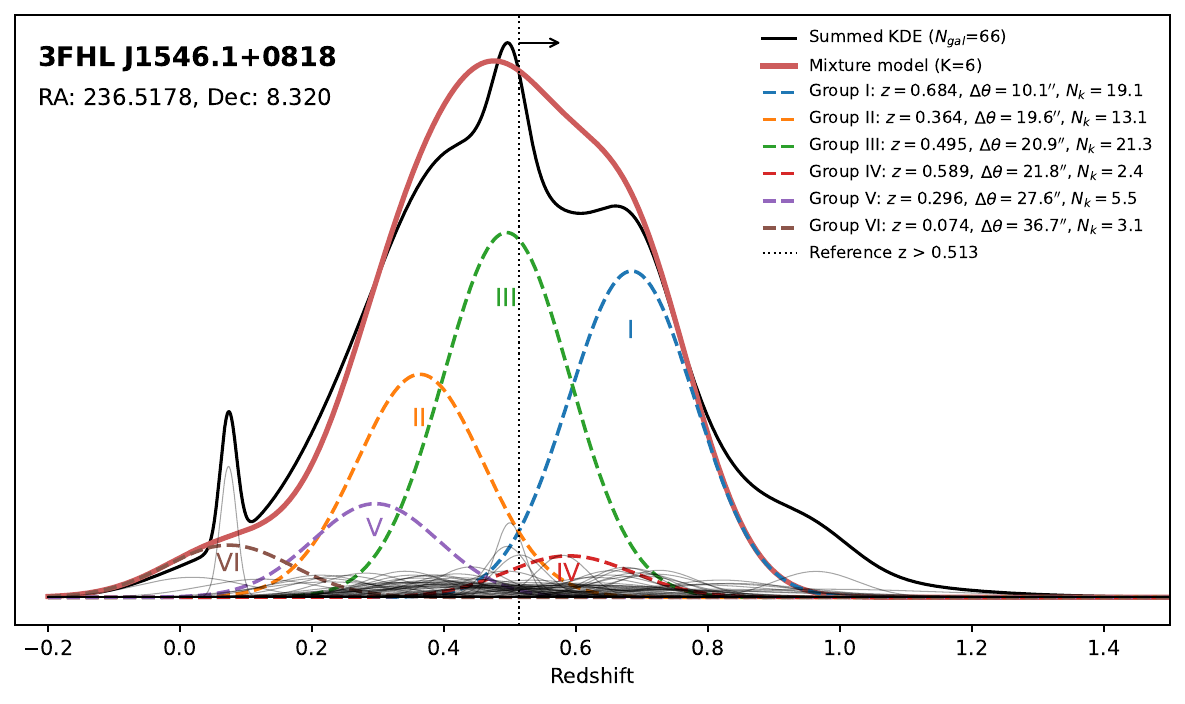}
  \includegraphics[width=0.49\textwidth]{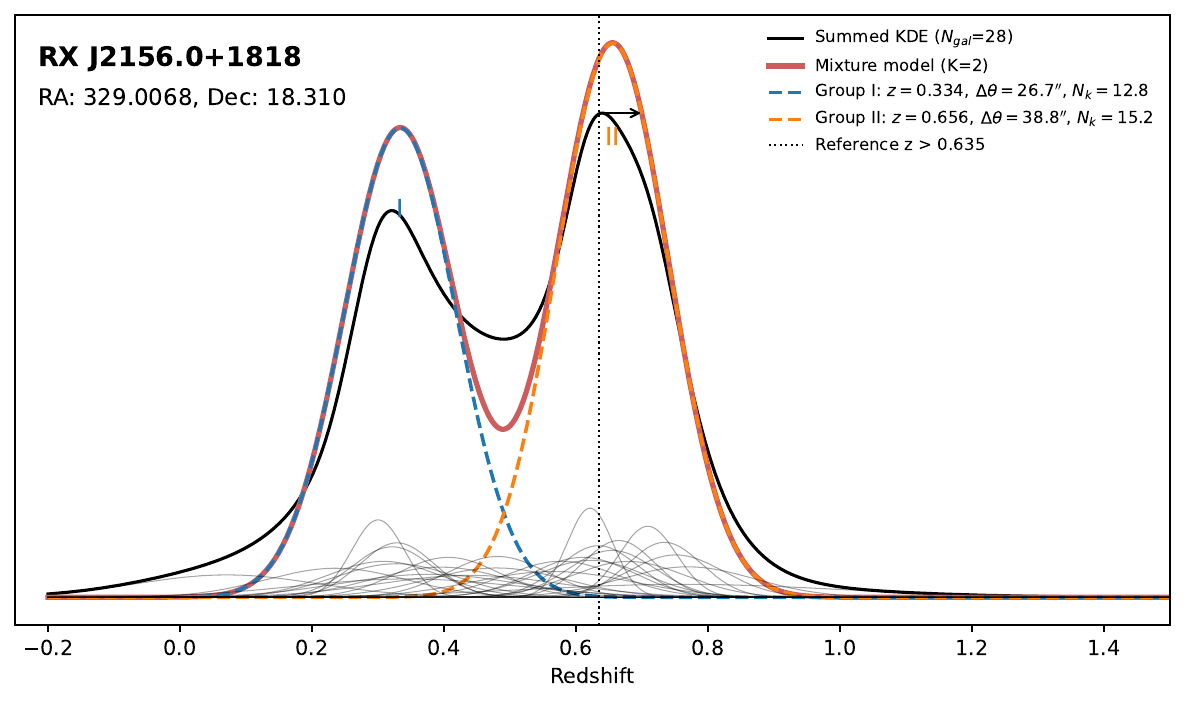}
  \caption{Redshift distribution of the SDSS galaxies in the fields of PKS~1413$+$135, 3FHL~J1447.9$+$3608, 3FHL~J1546.1$+$0818, and RX~J2156.0$+$1818. See Figure \ref{fig:pks1424} and Section \ref{sec:method} for explanation of the data.} \label{fig:blazar_res3}
\end{figure*}

We now apply our method to blazars whose redshift information is uncertain or available only as limits. The results are summarised in Table~\ref{tab:blazar_res} and shown in Figs.~\ref{fig:blazar_res1}, \ref{fig:blazar_res2}, and \ref{fig:blazar_res3}.

\begin{table*}
    \caption{Blazars with uncertain or unknown redshifts within the SDSS footprint. Columns: (1) blazar identifier; (2) reference for the published redshift estimate or limit; (3) literature-reported redshift estimate or limit; (4) redshift corresponding to the centroid of the closest GMM component to the spectroscopic redshift of the blazar (from our analysis); (5) difference between the group redshift and the literature value; (6) angular separation between the blazar and the composite central position of the corresponding GMM component; (7) rank order of the group composite centroid in the field (I = closest to the blazar); (8) alternative group redshift that could plausibly be associated with the blazar; (9) angular separation between the blazar and the alternative composite central position of the corresponding GMM component; (10) rank order of the alternative group composite centroid in the field.} \label{tab:blazar_res}
    \centering
    \begin{tabular}{l|ccccccccc}   
    \hline
    Source & Ref & z & Group-z & $\Delta$z & $\Delta\theta$('') & Group order  & alt. Group-z & alt. $\Delta\theta$('') & alt. Group order\\
    \hline
    TXS~0141$+$268 & 1 & $>$0.708 & 0.93 & -- & 29 & IV \\
    PKS~0735$+$178 & 2 & $>$0.42, $0.45\pm0.06$ & 0.39 & -0.03 & 20 & III & 0.62 & 8 & I \\ 
    3FHL~J0905.5$+$1357 & 3 & 0.64?, $0.50\pm0.09$ & 0.61 & -0.03 & 24 & III \\
    B2~0912$+$29 & 4 & $>0.13$, $0.36^{+0.07}_{-0.06}$, 0.53 & 0.57 & +0.04 & 9 & I \\
    GB6~J1058$+$2817 & 5 & 0.48? & 0.44 & -0.04 & 26 & IV & 0.59 & 18 & I   \\
    3FHL~J1120.8$+$4212 & 6 & $>$0.28 & 0.39 & -- & 6 & I \\
    3FHL~J1150.5$+$4154 & 7 &$>$0.25, 0.46$\pm$0.09 & 0.39 & -0.07 & 5 & II & 0.55 & 13 & III \\ 
    BZB~J1243$+$3627 & 8 & $>$0.48 & 0.47 & -0.01 & 13 & I \\
    3FHL~J1253.1$+$5300 & 9 & $>$0.66 & 0.69 & +0.03 & 7 & I \\
    PKS~1413$+$135 & 10 & 0.25$<z<$0.5 & 0.44 & -- & 7 & II & 0.36 & 13 & III \\ 
    3FHL~J1447.9$+$3608 & 11 & $>$0.74 & 0.80 & +0.06 & 10 & I \\
    3FHL~J1546.1$+$0818 & 12 & $>$0.513 & 0.50 & -0.01 & 21 & III & 0.68 & 10 & I \\
    RX~J2156.03$+$1818 & 13 & $>$0.63, 0.60$\pm$0.11 & 0.66 & -- & 15 & II \\
    \hline
    \end{tabular}\\
    \textbf{Redshift methods and references}:  
    1) Spectroscopy, intervening absorber \citep{shaw13}, 
    2) Spectroscopy, intervening absorber \citep{carswell74, mishra18}; host galaxy imaging \citep{nilsson12}, 
    3) Tentative, based on single spectral line \citep{2020MNRAS.497...94P}; host galaxy imaging \citep{nilsson24}, 
    4) Non-detection of host galaxy features in high S/N spectra \citep{paiano17}; host galaxy imaging \citep{meisner10}; weak-lensing cluster \citep{2006ApJ...643..128W},
    5) Tentative, based on single spectral line \citep{massaro14}, 
    6) Non-detection of host galaxy features in high S/N spectra \citep{paiano17},
    7) Non-detection of host galaxy features in high S/N spectra \citep{2020MNRAS.497...94P}; host galaxy imaging \citep{nilsson24},
    8) Spectroscopy, intervening absorber \citep{plotkin10,shaw13,mishra18}, 
    9) Spectroscopy, intervening absorber \citep{shaw13,mishra18,2020MNRAS.497...94P}, 
    10) Millilensing arguments \citep{readhead21}, 
    11) Spectroscopy, intervening absorbed \citep{shaw13,mishra18,2020MNRAS.497...94P}, 
    12) Spectroscopy, intervening absorber \citep{ahumada20}, 
    13) Spectroscopy, intervening absorber \citep{dammando24}; host galaxy imaging \citep{nilsson24}.
\end{table*}    

For each source we identify several candidate galaxy structures (GMM components) along the line of sight. As a baseline association criterion, we adopt the component whose composite centroid lies closest to the blazar position on the sky (smallest $\Delta\theta$). When independent redshift constraints exist (e.g. intervening absorption systems, tentative emission-line identifications, host-galaxy imaging, or $\gamma$-ray-based limits), we use them to assess whether the closest-centroid component is physically plausible. If the closest-centroid component is inconsistent with robust external constraints, we instead report an alternative component that satisfies those constraints and list it as an `alternative' association in Table~\ref{tab:blazar_res}. In ambiguous cases, we refrain from selecting a unique redshift and discuss the viable options in Section~\ref{sec:notes}.

We caution that the SDSS-based group associations become incomplete at both low and high redshift for different reasons. At low redshift, the fixed $2\arcmin$ aperture corresponds to a smaller physical radius and may not sample the full spatial extent of a group, biasing the inferred centroid. At high redshift ($z \gtrsim 0.8$), SDSS photometry becomes increasingly incomplete for typical group members, so some genuine overdensities may not be recovered, and fitted high-$z$ components may be supported by only a small number of galaxies. These limitations are reflected in the larger centroid offsets and the increased frequency of multiple plausible components in some fields.

\subsection{Notes on individual sources} \label{sec:notes}

\textbf{TXS~0141$+$268:} \citet{shaw13} set a lower limit of $z>0.708$ based on an intervening absorber. The machine-learning approach of \citet{narendra22}, based on \emph{Fermi}-LAT data, yields a redshift estimate of either $z=0.73$ or $z=0.95$ (bias-corrected). While the accuracy of these values is uncertain, they indicate a relatively large distance and are consistent with the lower limit from \citet{shaw13}. Above this limit, we identify two candidate groups at $z=0.93$ (fourth closest centroid; $\Delta\theta=29\arcsec$) and at $z=0.83$ (sixth closest centroid; $\Delta\theta=48\arcsec$). Although these components consist of only a few SDSS photometric redshifts (owing to the SDSS magnitude limit), the component at $z=0.93$ is consistent with the $\gamma$-ray-based estimate supports the association. \textbf{Summary: Likely redshift $z\sim0.93-0.95$}.

\textbf{PKS~0735$+$178:} A lower limit of $z=0.424$ was determined by \citet{carswell74} from a Mg\,\textsc{ii} absorption doublet and later confirmed by \citet{mishra18}. From host galaxy imaging, \citet{nilsson12} estimated $z=0.45\pm0.06$, consistent with the absorption system if BL~Lac hosts are treated as standard candles. \citet{acharyya23} argued that this value might be underestimated and that $z\sim0.8$ could better explain VHE upper limits when extrapolating the \emph{Fermi}-LAT spectrum (though alternative interpretations, such as an intrinsic spectral cutoff near 100\,GeV or strong internal absorption are also possible). We decompose the SDSS field into four candidate structures at $z=0.19$, $z=0.39$, $z=0.62$, and $z=0.82$. The $z=0.39$ component agrees with both the absorption and imaging estimates but has a relatively large offset ($\Delta\theta=20\arcsec$). The $z=0.82$ component is compatible with the $\gamma$-ray-based redshift estimate but it has the largest centroid offset in the field ($\Delta\theta=28\arcsec$). The closest centroid lies at $z=0.62$ ($\Delta\theta=8\arcsec$). The nearest companion galaxy, $7\arcsec$ northwest of the blazar, has a spectroscopic redshift of $z=0.645$ \citep{stickel93}; SDSS spectroscopy reveals another galaxy at a similar redshift (Appendix~\ref{sec:sdss_spec}, Table~\ref{tab:sdss_redshift}), consistent with previous reports \citet{2021ATel15132....1F}. If the blazar is associated with a group, the most likely redshift is therefore $z\approx0.64$. This would imply a luminous host galaxy; however, as discussed by \citet{nilsson24}, at higher redshifts only the brightest hosts are expected to be detectable. \textbf{Summary: Likely redshift $z=0.64$–$0.65$}.  

\textbf{3FHL~J0905.5$+$1357:} Spectroscopy revealed a single emission line \citep{2020MNRAS.497...94P}. If identified as [O\,\textsc{iii}]$\lambda5007$, the implied redshift is $z=0.2239$; if identified as [O\,\textsc{ii}]$\lambda3727$, the redshift is $z=0.644$. Host-galaxy imaging by \citet{nilsson24} yielded $z=0.50\pm0.09$. The inferred absolute magnitudes would make the host relatively faint at $z=0.2239$ ($M_R=-20.9$), whereas at $z=0.644$ it would be luminous ($M_R=-23.4$) but still within the observed BL~Lac host distribution \citep[$M_R=-22.9$, $\sigma=0.5$;][]{sbarufatti05}, favouring the higher redshift. The machine-learning method of \citet{narendra22} gives $z=0.38$ (or $z=0.24$, bias-corrected), suggesting a closer distance. Our analysis identifies a component at $z=0.61$ (compatible with the $z=0.644$ interpretation), with a centroid offset $\Delta\theta=19\arcsec$. The five nearest SDSS galaxies within $30\arcsec$ have photometric redshifts consistent (within $\sim2\sigma$) with $z\approx0.64$, and SDSS includes at least one spectroscopic galaxy redshift close to $z=0.64$ in the immediate field (Appendix~\ref{sec:sdss_spec}, Table~\ref{tab:sdss_redshift}). \textbf{Summary: Likely redshift $z=0.644$}.

\textbf{B2~0912$+$29:} Several high-$S/N$ spectra are featureless \citep{shaw13,paiano17}. \citet{paiano17} derived a lower limit of $z>0.13$. \citet{meisner10} reported an image-based estimate of $z=0.36^{+0.07}_{-0.06}$, although \citet{nilsson24} could not confirm it due to anomalies in the image. The field was investigated in a weak-lensing shear cluster search by \citet{2006ApJ...643..128W}, who identified a cluster candidate DLCS~J0916.0$+$2931 at $z=0.53$, centred $2.5\arcmin$ from the blazar and extending $\sim10\arcmin$ north--south. We find a candidate component at $z=0.57$ that is the closest group in our analysis ($\Delta\theta=9\arcsec$) and consistent with the redshift of the weak-lensing cluster. \textbf{Summary: Likely redshift $z\approx0.53$}. 

\textbf{GB6~J1058$+$2817:} The literature redshift is tentative ($z=0.4793$; \citealt{massaro14}), based on an SDSS spectrum. We identify a component at $z=0.44$, consistent within the expected uncertainties, but with quite large centroid offset ($\Delta\theta=26\arcsec$). The closest component to the blazar in our analysis is located at $z=0.59$ ($\Delta\theta=18\arcsec$) \textbf{Summary: no unique redshift can be selected; plausible values are $z=0.48$ and $z\approx0.59$}.  

\textbf{3FHL~J1120.8$+$4212:} \citet{paiano17} reported a lower limit of $z>0.28$ from a featureless spectrum and considered earlier estimates unreliable. \citet{2022MNRAS.509.4330D} analysed archival \textit{HST}/COS G130M data and derived a constraint $z=0.201$–$0.228$ from the highest-redshift Ly$\alpha$ absorbed, noting that additional UV coverage is required to exclude higher-$z$ absorption unambiguously. In our analysis, the closest-centroid component lies at $z=0.39$, and SDSS spectroscopy yields eight galaxies clustered at $z\simeq0.363$ (Appendix~\ref{sec:sdss_spec}, Table~\ref{tab:sdss_redshift}). This strongly suggests an associated group at $z=0.363$. \textbf{Summary: likely redshift $z=0.363\pm0.002$}. 

\textbf{3FHL~J1150.5$+$4154:} Host-galaxy imaging gives $z=0.46\pm0.09$ \citep{nilsson24}, compatible (within $1\sigma$) with our candidate components at $z=0.39$ and $z=0.55$, where $z=0.39$ has a smaller centroid offset ($\Delta\theta=5\arcsec$). However, SDSS spectroscopy in the wider environment reveals a concentration of galaxies at $z\simeq0.323$ (Appendix~\ref{sec:sdss_spec}, Table~\ref{tab:sdss_redshift}), also discussed by \citet{nilsson24}. If the blazar were associated with this nearer structure, the detected host would be underluminous, but this cannot be excluded. \textbf{Summary: no unique redshift can be selected; a plausible range is $z\sim0.3$--0.55}.

\textbf{BZB~J1243$+$3627:} Mg\,\textsc{ii} absorption yields $z>0.48$ \citep{plotkin10,shaw13,mishra18}. This is consistent with the marginal host-galaxy detection of \citet{meisner10}, who estimated $z\sim0.50$. We identify three components close to and above the absorption limit: one at $z=0.47$ (closest centroid; $\Delta\theta=13\arcsec$), one at $z=0.67$ (second closest; $\Delta\theta=15\arcsec$) and one at $z=0.55$ (third closest; $\Delta\theta=17\arcsec$). Given the weak host detection and the proximity of the lower-$z$ component, we favour the $z=0.47$ association. SDSS also contains a nearby galaxy with spectroscopic redshift $z=0.485$ (Appendix~\ref{sec:sdss_spec}, Table~\ref{tab:sdss_redshift}), supporting a blazar redshift close to the absorber. \textbf{Summary: likely redshift $z\simeq0.485$}.

\textbf{3FHL~J1253.1$+$5300:} Mg\,\textsc{ii} absorption implies $z>0.66$ \citep{shaw13,mishra18,2020MNRAS.497...94P}. We recover a component at $z=0.69$ (the closest centroid; $\Delta\theta=7\arcsec$) and another at $z=0.89$ (fourth closest centroid; $\Delta\theta=36\arcsec$). Given the absorber constraint and the published value $z=0.6638$ \citep{2020MNRAS.497...94P}, we adopt $z\simeq0.664$ as the most likely redshift. \textbf{Summary: likely redshift $z=0.6638$}.

\textbf{PKS~1413$+$135:} This is a complex line-of-sight system involving an edge-on foreground spiral galaxy at $z=0.247$ \citep{stocke92} and a background AGN \citep{readhead21}, potentially affected by millilensing \citep{peirson22}. \citet{lamer99} argued for an upper limit $z<0.5$ for the background source based on the lack of multiple lensed images. Within this interval, we find two plausible components: $z=0.44$ (second closest centroid; $\Delta\theta=7\arcsec$) and $z=0.36$ (third closest; $\Delta\theta=13\arcsec$). Both satisfy the lensing-based upper limit and are viable associations. \textbf{Summary: no unique redshift can be selected; plausible values are $z\sim0.3$--0.5}.

\textbf{3FHL~J1447.9$+$3608:} Mg\,\textsc{ii} absorption implies $z>0.738$ \citep{shaw13,mishra18,2020MNRAS.497...94P}. We recover a component at $z=0.80$ (the closest centroid; $\Delta\theta=10\arcsec$), consistent with this absorber. However, the SDSS depth limits sensitivity to even higher-$z$ structures, so the true redshift could be larger. \textbf{Summary: no definitive redshift can be selected; $z\gtrsim0.74$ is supported}.

\textbf{3FHL~J1546.1$+$0818:} Mg\,\textsc{ii} absorption implies $z>0.513$ \citep{ahumada20}, and host-galaxy imaging also suggests $z\gtrsim0.5$ \citep{nilsson24}. We find a component near the absorber redshift at $z=0.50$ (third closest centroid; $\Delta\theta=21\arcsec$). The closest-centroid component lies at $z=0.68$ ($\Delta\theta=10\arcsec$) and is also consistent with the lower limits. \textbf{Summary: no unique redshift can be selected; plausible values are $z=0.5$ and $z\approx0.68$}.

\textbf{RX~J2156.0$+$1818:} Mg\,\textsc{ii} absorption yields $z>0.635$ \citep{dammando24}, while host-galaxy imaging gives $z=0.60\pm0.11$ \citep{nilsson24}. We identify a component at $z=0.66$, consistent with both constraints. No additional higher-$z$ components are detected in SDSS, suggesting that the redshift is not far above the absorber. \textbf{Summary: likely redshift $z\simeq0.635$}.

\section{Conclusions}

We have extended the cosmic-neighbour association method to estimate distances for BL~Lac objects with uncertain redshifts or only limit-based constraints. By identifying galaxy overdensities in SDSS fields using photometric redshifts and associating these structures with the blazar position, we obtain group-based redshift estimates over a wide range of distances ($z\sim0.2$--$0.9$). Tests on control samples with spectroscopic blazar redshifts and with MOS-studied environments show that the redshifts of associated galaxy structures are recovered with a typical accuracy of $\Delta z \sim 0.02$--0.04.

Applying the method to BL~Lacs lacking secure spectroscopic redshifts, we identify plausible cosmic-neighbour associations and provide refined distance constraints. For a subset of sources, consistent evidence from intervening absorption systems, host-galaxy imaging, $\gamma$-ray-based constraints, and the SDSS group-association analysis allows us to propose likely redshifts for TXS~0141$+$268 ($z\sim0.93-0.95$), PKS~0735$+$178 ($z\simeq0.64$), 3FHL~J0905.5$+$1357 ($z=0.644$), B2~0912$+$29 ($z=0.53$), 3FHL~J1120.8$+$4212 ($z=0.363$), BZB~J1243$+$3627 ($z=0.485$), 3FHL~J1253.1$+$5300 ($z=0.664$), and RX~J2156.0$+$1818 ($z=0.635$). For other objects, we identify multiple viable structures along the line of sight and outline the remaining degeneracies.

Our analysis demonstrates that environmental (or `cosmic neighbour') information provides significant leverage in constraining redshifts for BL~Lac objects, particularly those lacking measurable spectral features. The technique is especially valuable for VHE blazars, whose intrinsic properties and $\gamma$-ray attenuation depend sensitively on distance. Reliable redshift estimates are therefore essential for interpreting their spectral energy distributions and for testing extragalactic background light and intergalactic magnetic field models.

Nonetheless, certain caveats remain. At low redshift, the fixed $2\arcmin$ aperture corresponds to a smaller physical radius and can bias centroid-based association metrics. At high redshift ($z\gtrsim0.8$), the SDSS magnitude limit reduces the number of detected group members, so genuine overdensities may be missed and high-$z$ components may be supported by only a few galaxies. These effects can lead to ambiguous associations in some fields, and group-based distances should therefore be interpreted as probabilistic constraints rather than definitive measurements.

Despite these limitations, the present results significantly expand the set of VHE blazars with plausible redshift determinations. Future data sets will substantially strengthen this approach. Deep wide-field imaging from surveys such as the Vera C. Rubin Observatory LSST will increase the completeness of galaxy catalogues around blazars, while narrow-band photometric surveys such as PAUS and J-PAS will deliver near-spectroscopic photometric-redshift precision. Together, these improvements will enhance the detectability of group members, reduce line-of-sight projection ambiguities, and enable more reliable cosmic-neighbour distance estimates for BL~Lacs and other featureless AGN.

\section*{Acknowledgements}

We thank the anonymous referee for insightful comments on the statistical methods used in the paper.

This project has received funding from the European Research Council (ERC) under the European Union’s Horizon 2020 research and innovation programme (grant agreement No. 101002352, PI: M. Linares).

We acknowledge the financial support from the Visitor and Mobility program of the Finnish Centre for Astronomy with ESO (FINCA).

Funding for the Sloan Digital Sky Survey IV has been provided by the Alfred P. Sloan Foundation, the U.S. Department of Energy Office of Science, and the Participating Institutions. SDSS-IV acknowledges support and resources from the Center for High Performance Computing  at the University of Utah. SDSS-IV is managed by the Astrophysical Research Consortium for the Participating Institutions of the SDSS Collaboration including the Brazilian Participation Group, the Carnegie Institution for Science, Carnegie Mellon University, Center for Astrophysics | Harvard \& Smithsonian, the Chilean Participation Group, the French Participation Group, Instituto de Astrof\'isica de Canarias, The Johns Hopkins University, Kavli Institute for the Physics and Mathematics of the Universe (IPMU) / University of Tokyo, the Korean Participation Group, Lawrence Berkeley National Laboratory, Leibniz Institut f\"ur Astrophysik Potsdam (AIP),  Max-Planck-Institut f\"ur Astronomie (MPIA Heidelberg), Max-Planck-Institut f\"ur Astrophysik (MPA Garching), Max-Planck-Institut f\"ur Extraterrestrische Physik (MPE), National Astronomical Observatories of China, New Mexico State University, New York University, University of Notre Dame, Observat\'ario Nacional / MCTI, The Ohio State University, Pennsylvania State University, Shanghai Astronomical Observatory, United Kingdom Participation Group, Universidad Nacional Aut\'onoma de M\'exico, University of Arizona, University of Colorado Boulder, University of Oxford, University of Portsmouth, University of Utah, University of Virginia, University of Washington, University of Wisconsin, Vanderbilt University, and Yale University.

This work has made use of CosmoHub, developed by PIC (maintained by IFAE and CIEMAT) in collaboration with ICE-CSIC. It received funding from the Spanish government (grant EQC2021-007479-P funded by MCIN/AEI/10.13039/501100011033), the EU NextGeneration/PRTR (PRTR-C17.I1), and the Generalitat de Catalunya.

\section*{Data Availability}

The SDSS data are available at the SDSS website: \url{https://www.sdss4.org}. PAUS and J-PAS data are available at their respective websites: \url{https://pausurvey.org/public-data-release/} and \url{https://www.j-pas.org/datareleases/jpas_early_data_release}.



\bibliographystyle{mnras}
\bibliography{redshift} 




\appendix

\section{Mixture model and the expectation--maximisation algorithm}
\label{app:em}

Here, we summarise the mixture model and the EM updates used to infer redshift structures from SDSS photometric redshifts.

From above, the analytic error-convolved likelihood is
\begin{equation}
p(z_i)=\sum_{k=1}^{K_{\max}} w_k\,\mathcal{N}\!\left(z_i \mid \mu_k,\, \sigma_{z,i}^2\right),
\end{equation}
and the corresponding log-likelihood
\begin{equation}
\ln\mathcal{L} = \sum_i \ln\left[\sum_{k=1}^{K_{\max}} w_k\,\mathcal{N}\!\left(z_i \mid \mu_k,\, \sigma_{z,i}^2\right)\right].
\end{equation}

EM proceeds by alternating between
\emph{i)} an E-step computing posterior membership probabilities
\begin{equation}
r_{ik}=
\frac{
w_k\,\mathcal{N}\!\left(z_i \mid \mu_k,\, \sigma_{z,i}^2\right)
}{
\sum_{j=1}^{K_{\max}} w_j\,\mathcal{N}\!\left(z_i \mid \mu_j,\, \sigma_{z,i}^2\right)
},
\end{equation}
and \emph{ii)} an M-step updating weights and means:
\begin{equation}
w_k=\frac{1}{N}\sum_i r_{ik},
\qquad
\mu_k=\frac{\sum_i r_{ik}\,z_i/\sigma_{z,i}^2}{\sum_i r_{ik}/\sigma_{z,i}^2}.
\end{equation}
We iterate until the fractional change in $\ln\mathcal{L}$ falls below a tolerance ($10^{-6}$) or a maximum number of iterations ($n_{\max}=1000$) is reached.

Because EM can converge to local maxima, we adopt a multi-start strategy. For each field, we initialise $\mu_k$ by drawing $K_{\max}$ galaxies at random from the observed $z_i$ values (without replacement) and set $w_k=1/K_{\max}$. We repeat the EM fit for multiple random initialisations ($n_{\rm init}=100$) and retain the solution with the highest likelihood.

\section{SDSS redshift measurements from spectroscopy}
\label{sec:sdss_spec}

In addition to SDSS photometric redshifts, a small number of galaxies in several fields have SDSS spectroscopic redshifts. These measurements provide an internal consistency check on the redshift structures identified by our photometric-redshift GMM procedure and, in some cases, directly confirm the presence of a galaxy overdensity close to the redshift inferred for the blazar.

Table~\ref{tab:sdss_redshift} lists, for each blazar field analysed in this work, the number of galaxies within our $2\arcmin$ search radius with SDSS photometric redshifts ($N_{\rm photo\text{-}z}$) and with SDSS spectroscopic redshifts ($N_{\rm spectro\text{-}z}$), together with the full set of SDSS spectroscopic redshifts present in the field. Values are reported as provided by the SDSS pipeline; the number in parentheses indicates the uncertainty on the last quoted digit(s).

\begin{table*}
    \caption{Summary of SDSS redshift information within $2\arcmin$ of each blazar. Columns: (1) field (blazar) name; (2) number of galaxies with SDSS photometric redshifts used in the group analysis; (3) number of galaxies with SDSS spectroscopic redshifts in the same field; (4) list of all SDSS spectroscopic redshifts in the field, where parentheses indicate the uncertainty on the last digit(s) as reported by SDSS.} \label{tab:sdss_redshift}
    \centering
    \begin{tabular}{l|ccc}   
    \hline
    Field & $N_{\rm photo-z}$ & $N_{\rm spectro-z}$ & All SDSS spectroscopic redshifts in the field \\
    \hline
    RGB~J0013$+$191  & 37 & 1 & 0.51590(7) \\
    RGB~J0115$+$253  & 61 & 5 & 0.18232(9), 0.18417(3), 0.18631(2), 0.5790(3), 0.80598(3) \\  
    PKS~0139$-$09    & 54 & 0 \\
    RGB~J0202$+$088  & 46 & 0 \\
    RGB~J0227$+$020  & 104 & 4 & 0.18855(2), 0.4541(1), 0.4545(2), 0.90615(6) \\     
    GB6~J0154$+$0823 & 30 & 1 & 0.7009(2) \\
    IVS~B0200$+$30A  & 23 & 1 & 0.7768(3) \\
    RGB~J0757$+$099  & 47 & 0 \\
    1ES~0806$+$524   & 41 & 1 & 0.7155(2) \\
    PKS~0823$+$033   & 78 & 1 & 0.5043(1) \\
    1ES~1011$+$496   & 69 & 3 & 0.15507(3), 0.15522(1), 0.3796(1) \\
    ON~325           & 73 & 3 & 0.2791(2), 0.39448(8), 0.6464(3) \\
    W~Comae          & 70 & 2 & 0.10245(3), 0.10310(2) \\
    RGB~J1415$+$485  & 62 & 0 \\
    OX~183           & 39 & 0 \\
    CTD~135          & 50 & 2 & 0.44070(7), 0.4446(1)\\    
    \hline
    TXS~0141$+$268 & 36 & 1 & 0.19057(2) \\
    PKS~0735$+$178 & 68 & 3 & 0.5766(2), 0.6282(2), 0.64650(6) \\ 
    3FHL~J0905.5$+$1357 & 66 & 1 & 0.6385(2) \\    
    GB6~J1058$+$2817 & 36 & 1 & 0.24107(2) \\
    3FHL~J1120.8$+$4212 & 95 & 8 & 0.3604(1), 0.3615(1), 0.36295(9), 0.36326(8), 0.3636(1), 0.36415(9), 0.3642(1), 0.3667(1) \\
    3FHL~J1150.5$+$4154 & 91 & 6 & 0.3217(1), 0.3219(1), 0.32501(7), 0.3268(1), 0.3272(1), 0.5970(2) \\ 
    BZB~J1243$+$3627 & 37 & 1 & 0.4851(1) \\
    3FHL~J1253.1$+$5300 & 52 & 0 \\
    PKS~1413$+$135 & 61 & 2 & 0.24676(5), 0.36860(8) \\ 
    3FHL~J1447.9$+$3608 & 49 & 1 & 0.54867(7) \\
    3FHL~J1546.1$+$0818 & 65 & 2 & 0.07034(2), 0.5130(1) \\
    RX~J2156.03$+$1818 & 28 & 0\\
    \hline
    \end{tabular}
\end{table*}    

\section{Control blazar KDE fits} \label{sec:control_plots}

Here we present the diagnostic plots for the control sample of BL~Lacs with spectroscopic redshifts (Table~\ref{tab:bcontrol}). These figures show, for each field, the SDSS photometric-redshift PDFs of galaxies within $2\arcmin$, the summed redshift distribution, and the best-fit Gaussian mixture model used to identify candidate redshift structures along the line of sight. The purpose of these plots is to visualise the performance of the method in cases where the blazar redshift is known independently.

\begin{figure*}
  \centering
  \includegraphics[width=0.49\textwidth]{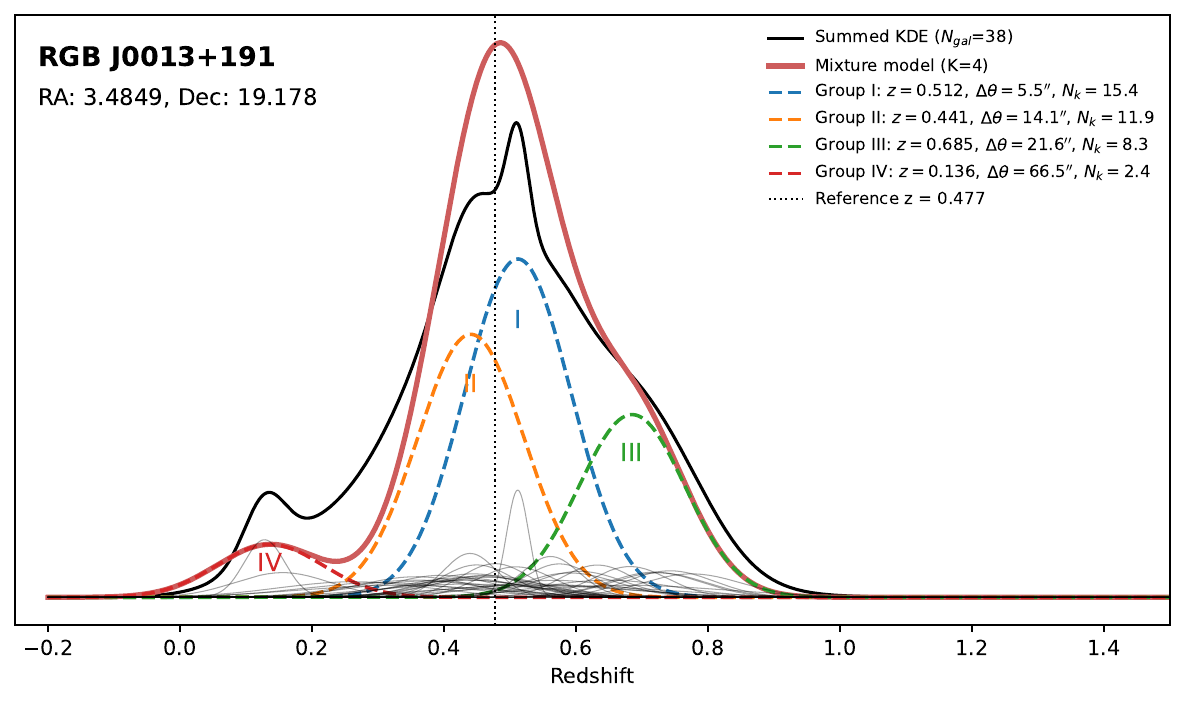}
  \includegraphics[width=0.49\textwidth]{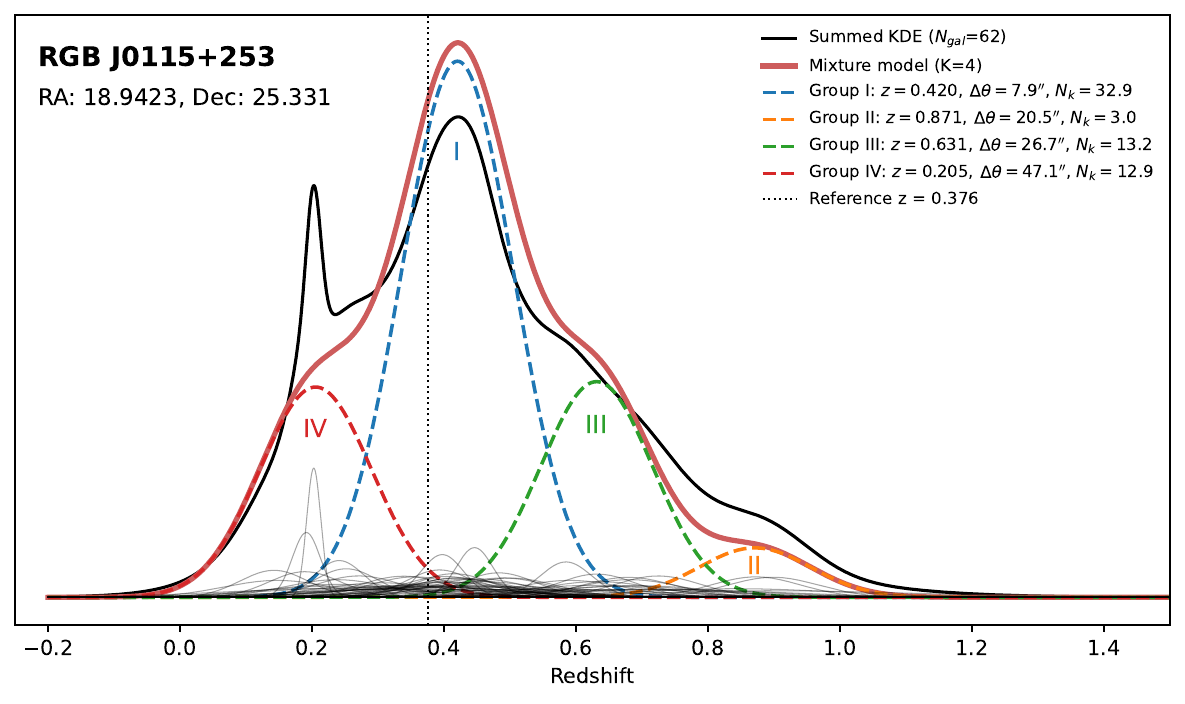}  
  \includegraphics[width=0.49\textwidth]{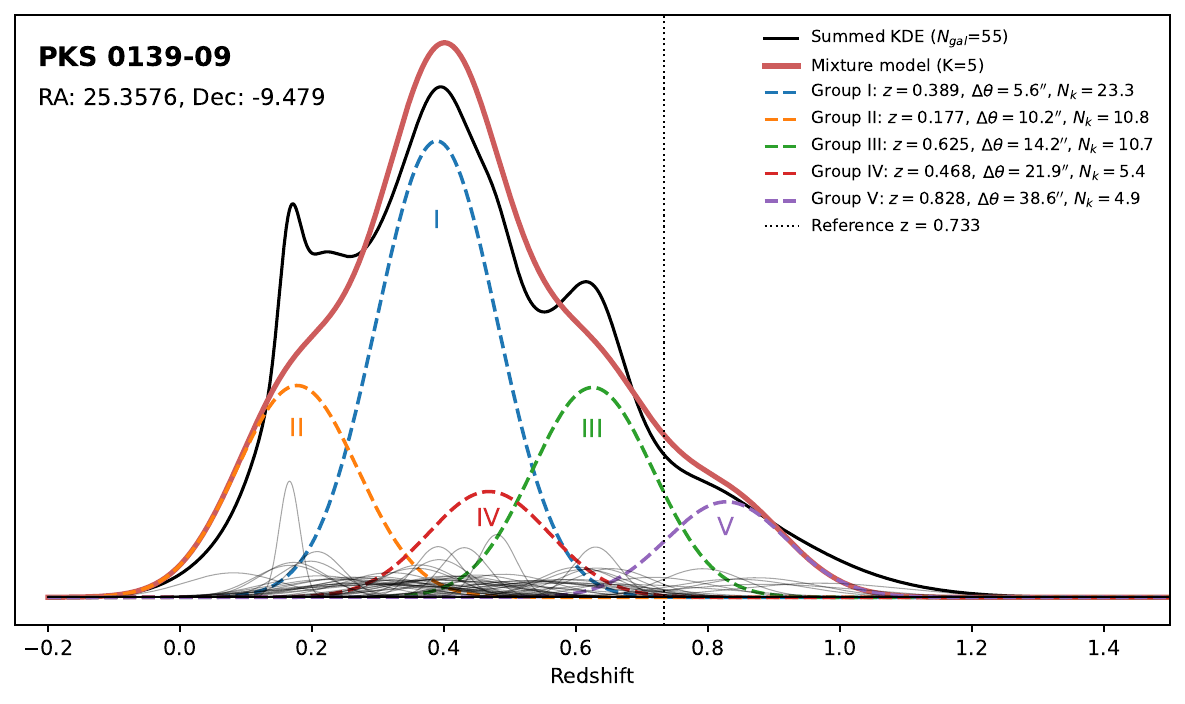}
  \includegraphics[width=0.49\textwidth]{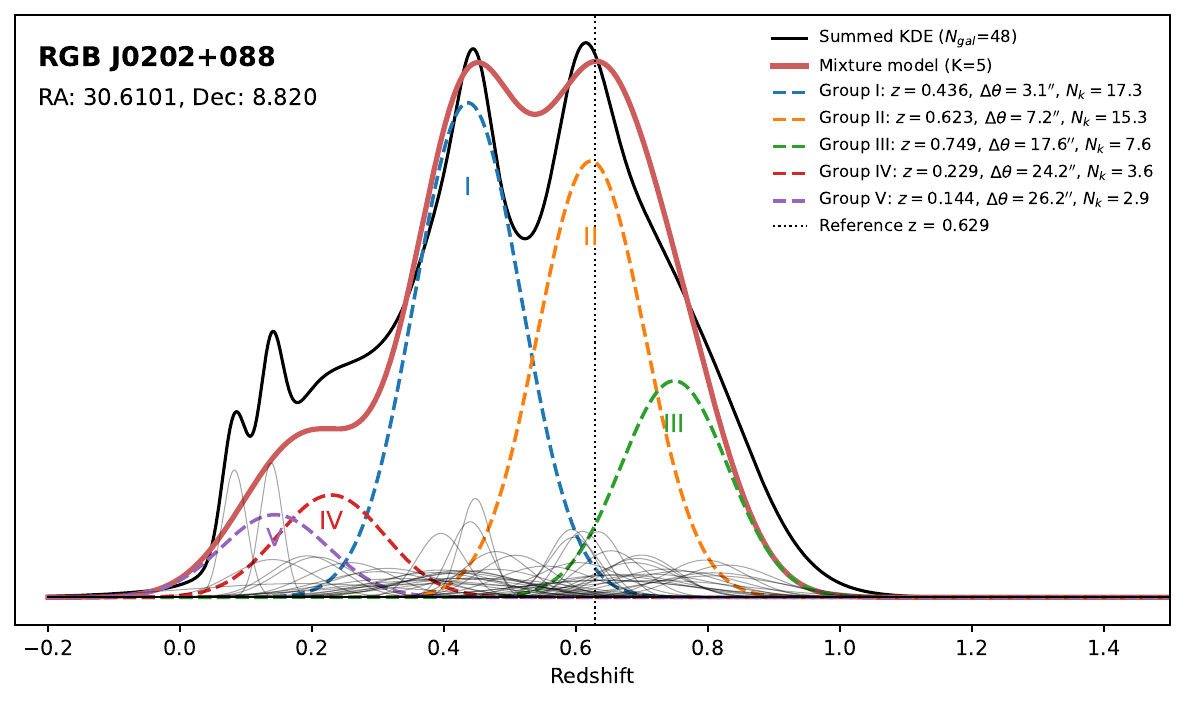}
  \includegraphics[width=0.49\textwidth]{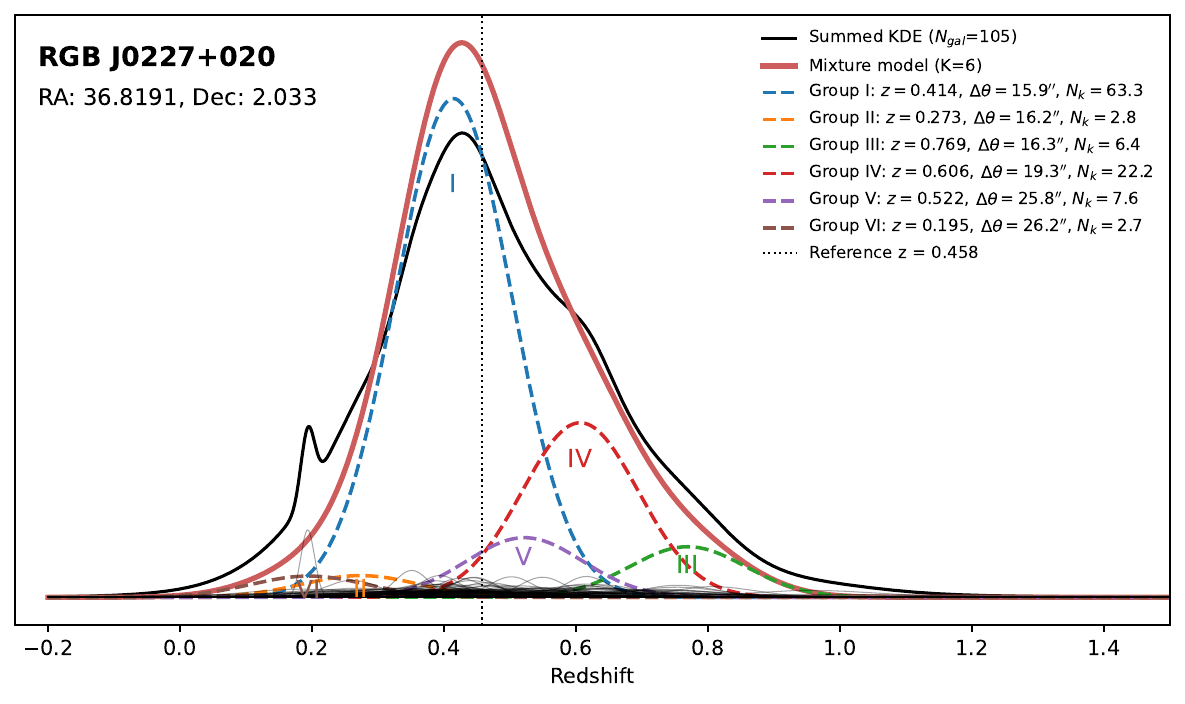}  
  \includegraphics[width=0.49\textwidth]{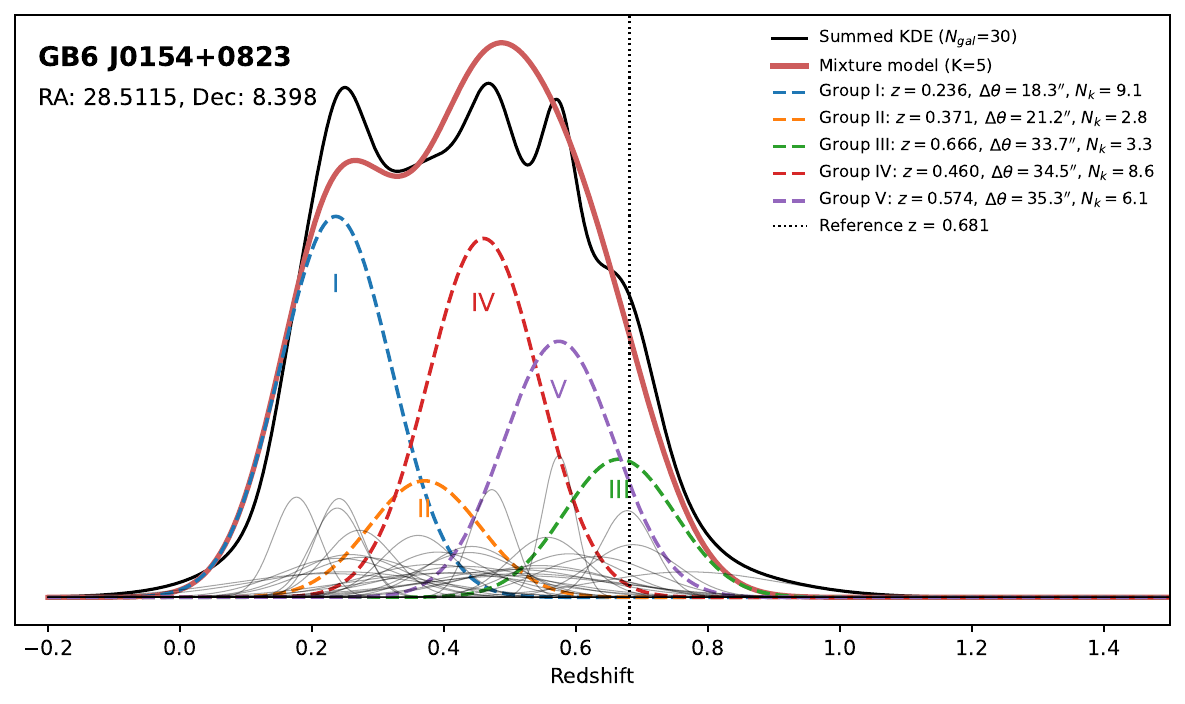}
  \includegraphics[width=0.49\textwidth]{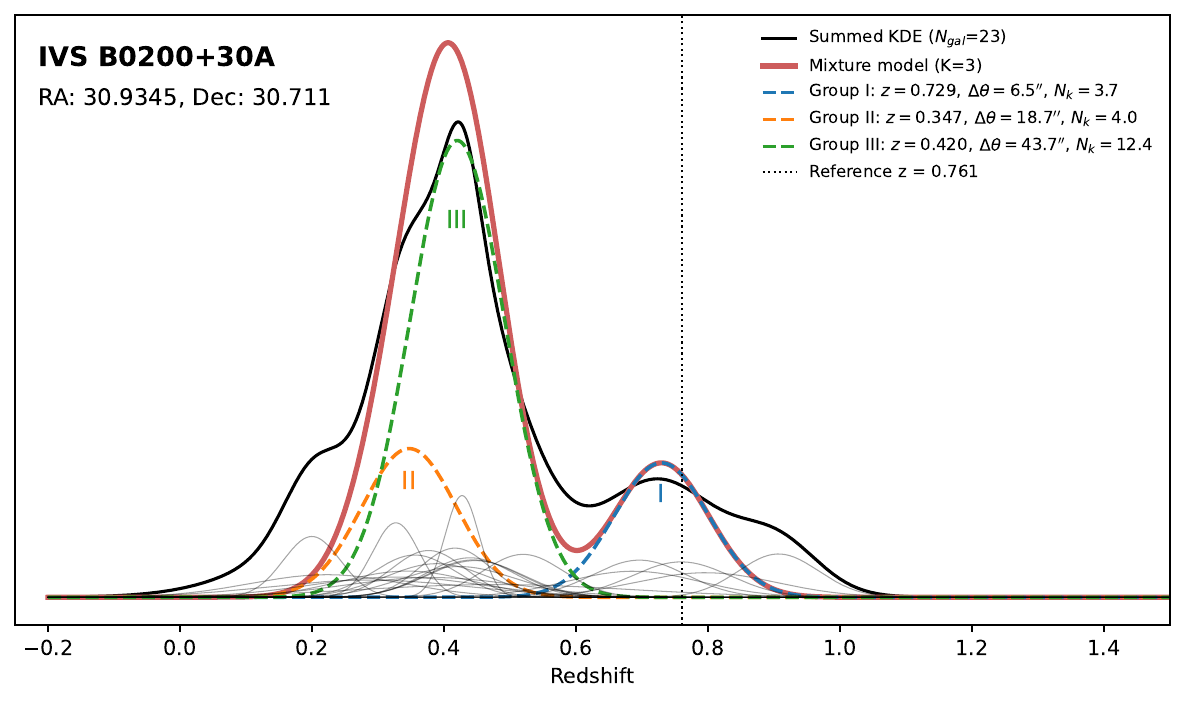}
  \includegraphics[width=0.49\textwidth]{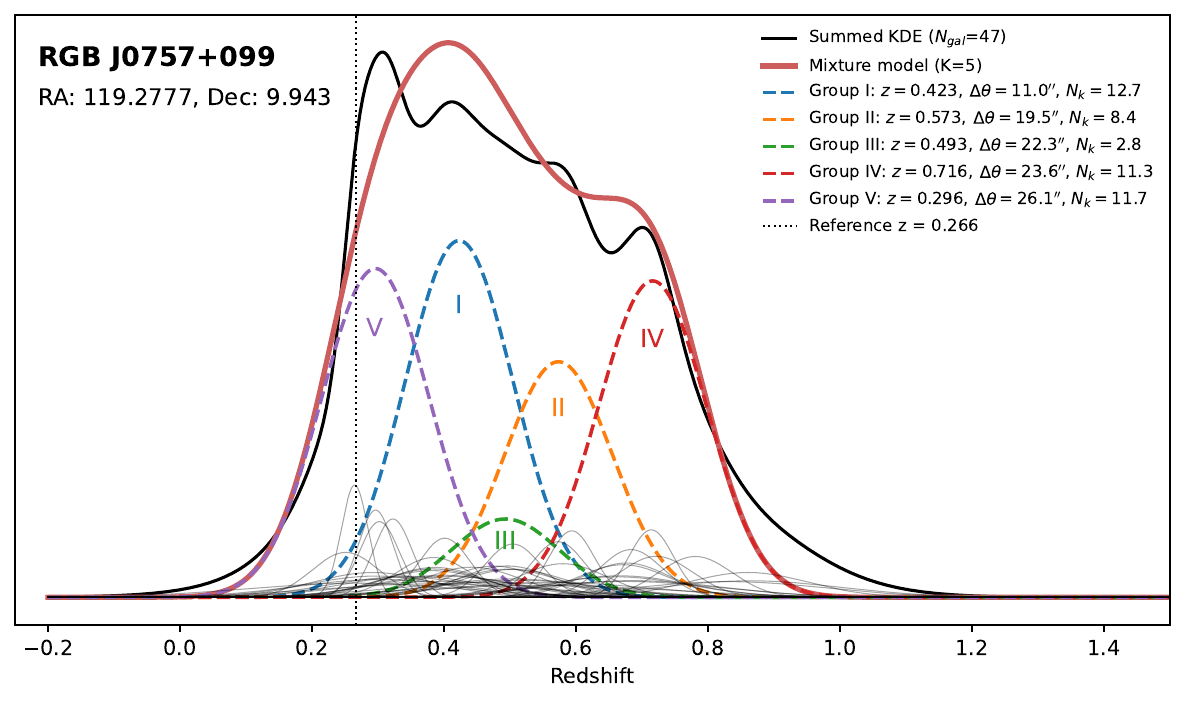} 
  \caption{Diagnostic plots (Section~\ref{sec:method}) showing the redshift distribution of SDSS galaxies within $2\arcmin$ of the control blazars RGB~J0013$+$191, RGB~J0115$+$253, PKS~0139$-$09, RGB~J0202$+$088, RGB~J0227$+$020, GB6~J0154$+$0823, IVS~B0200$+$30A, and RGB~J0757$+$099. See Fig.~\ref{fig:pks1424} and Section~\ref{sec:method} for a description of the plot elements.} \label{fig:control_spec1}
\end{figure*}

\begin{figure*}
  \centering
  \includegraphics[width=0.49\textwidth]{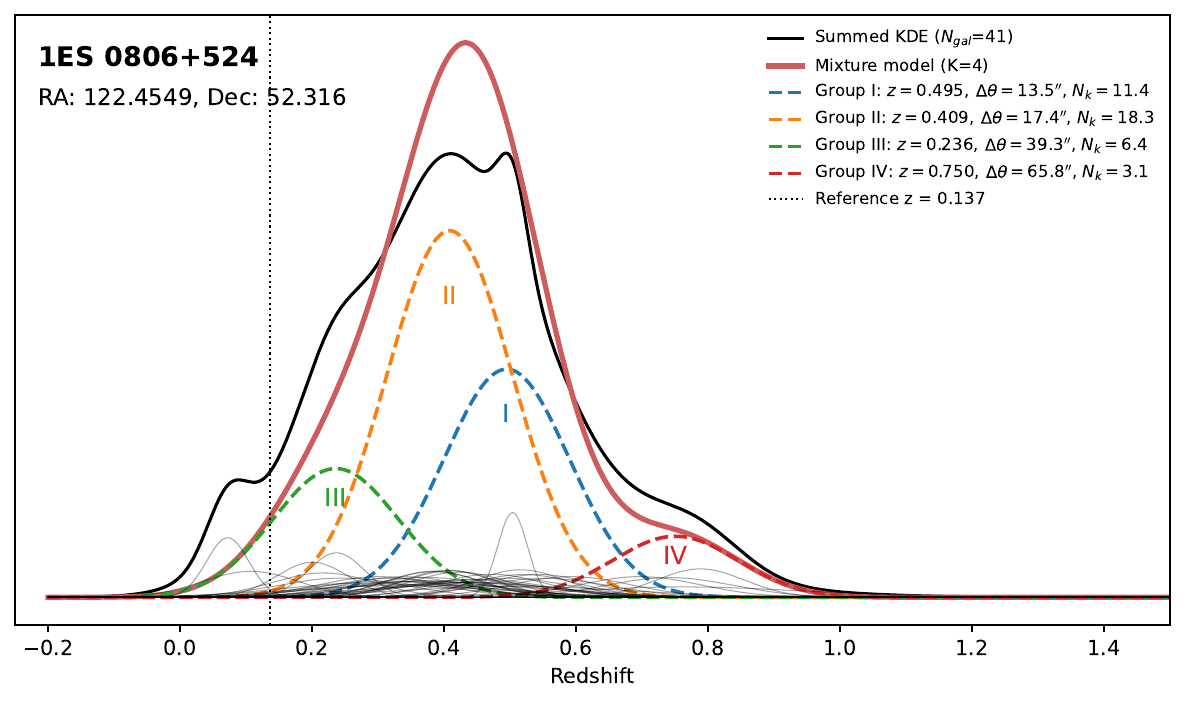}
  \includegraphics[width=0.49\textwidth]{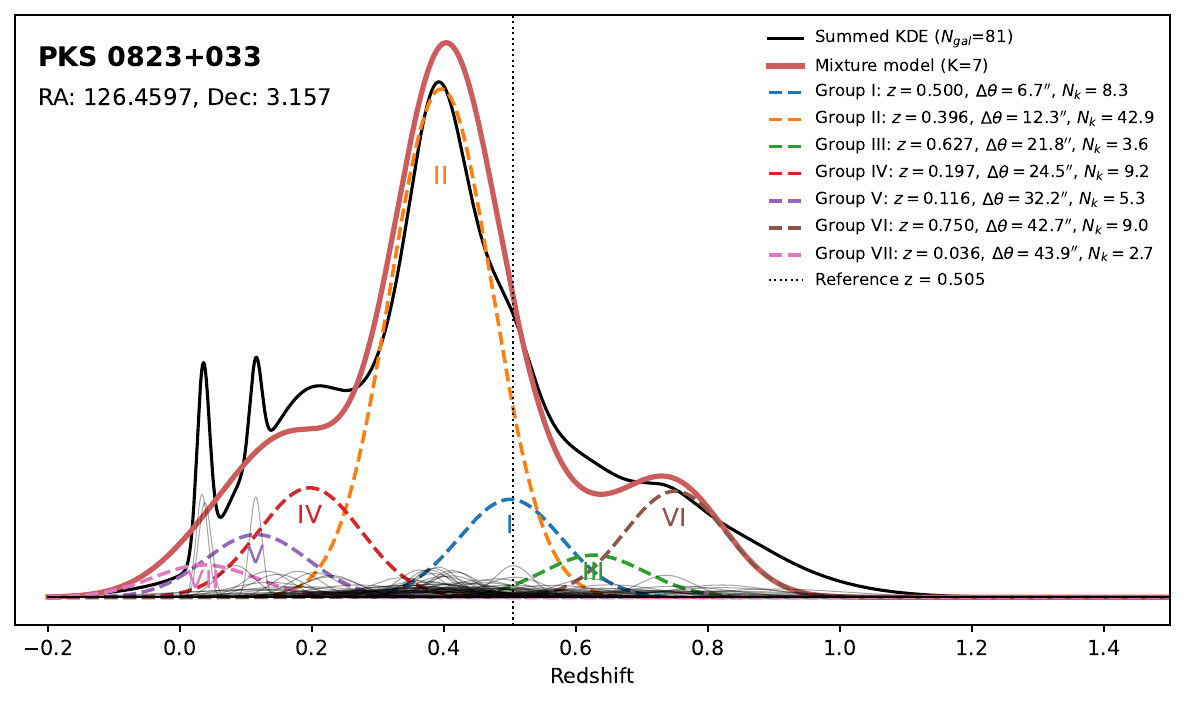}  
  \includegraphics[width=0.49\textwidth]{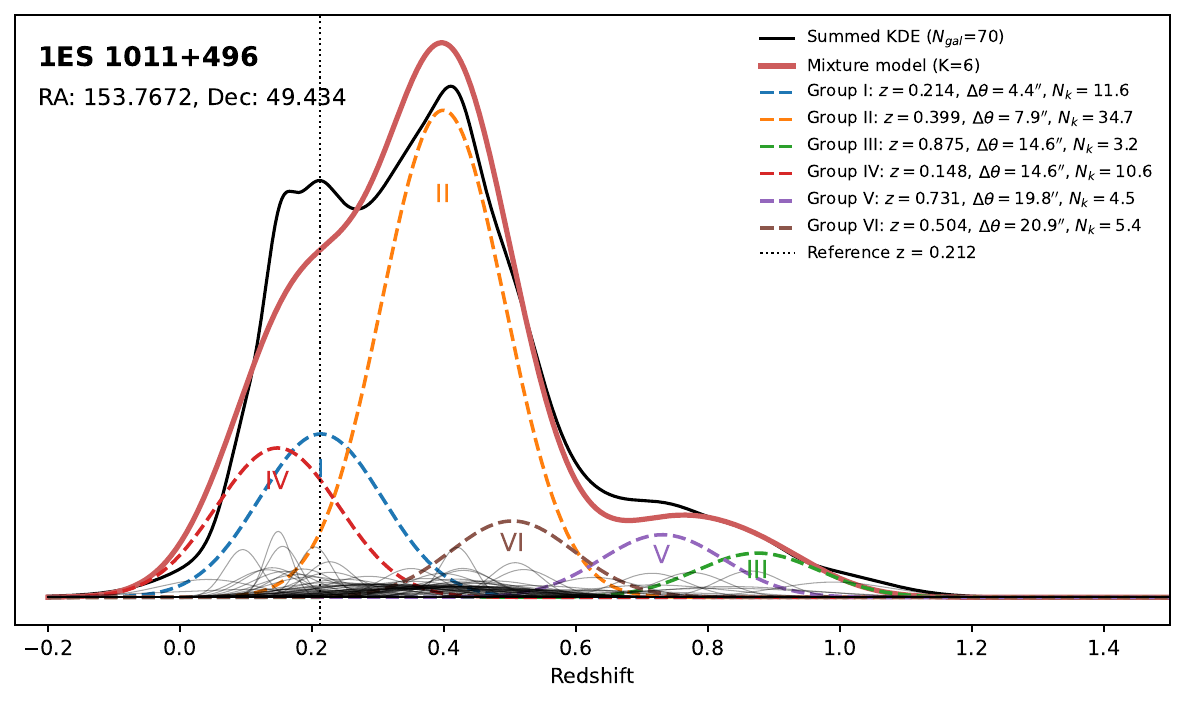}
  \includegraphics[width=0.49\textwidth]{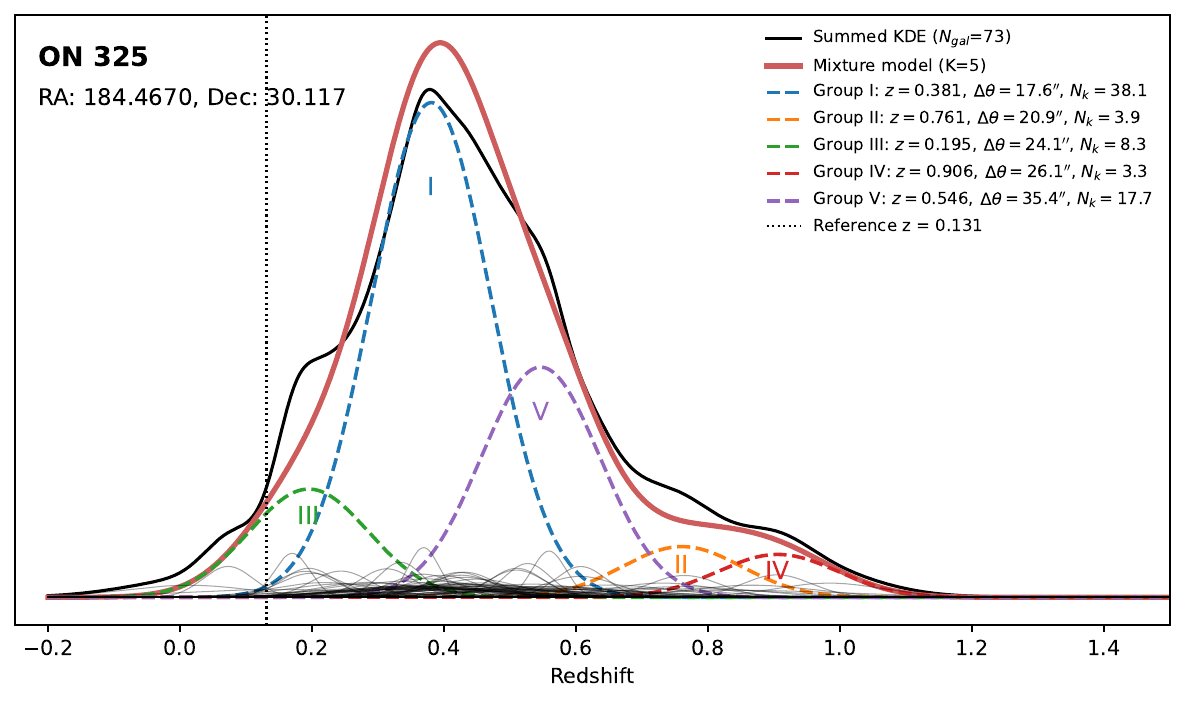}  
  \includegraphics[width=0.49\textwidth]{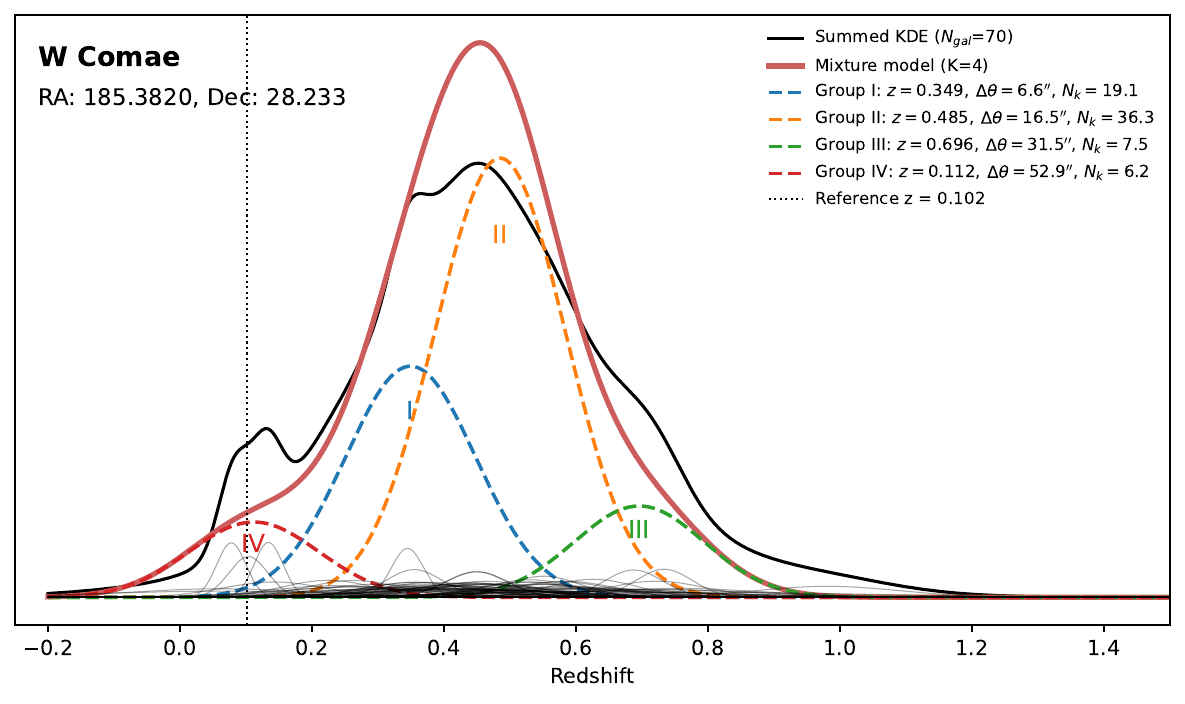}
  \includegraphics[width=0.49\textwidth]{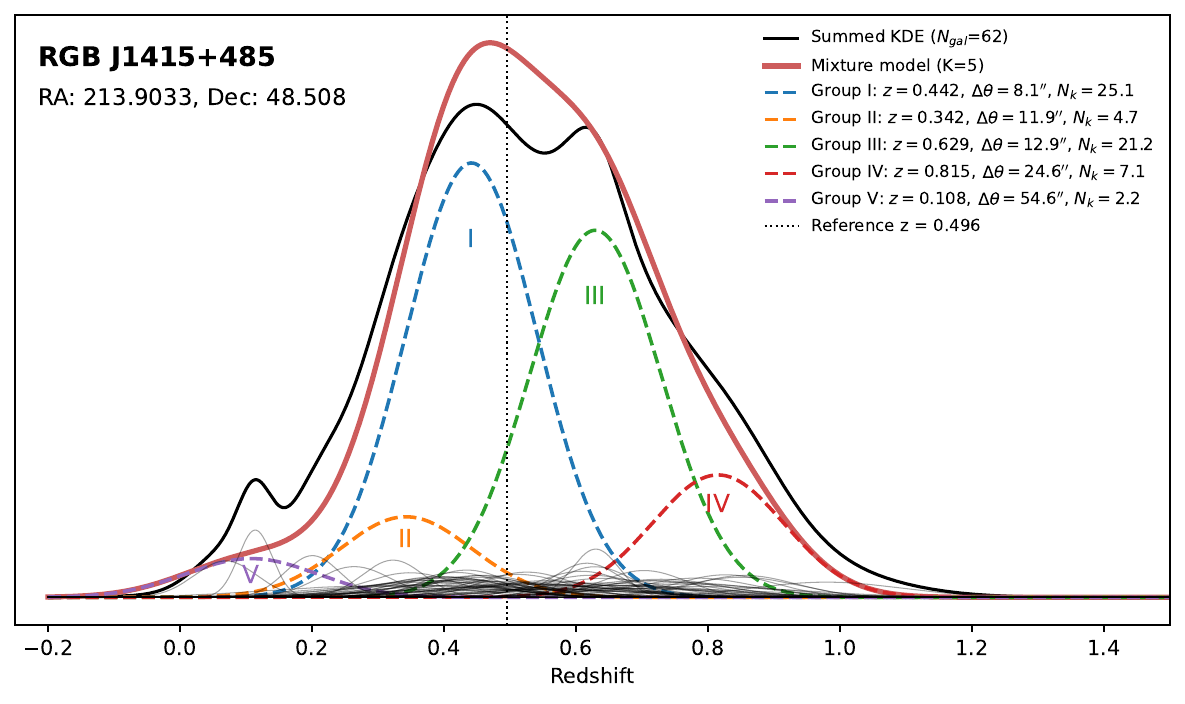} 
  \includegraphics[width=0.49\textwidth]{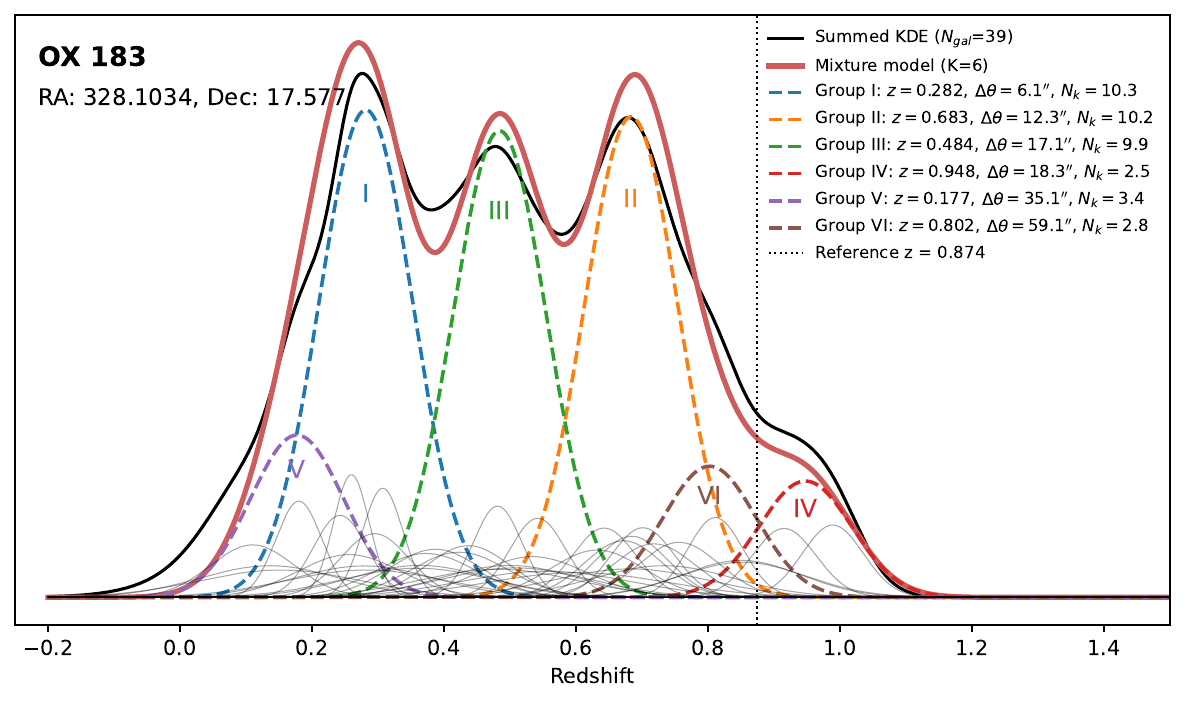}
  \includegraphics[width=0.49\textwidth]{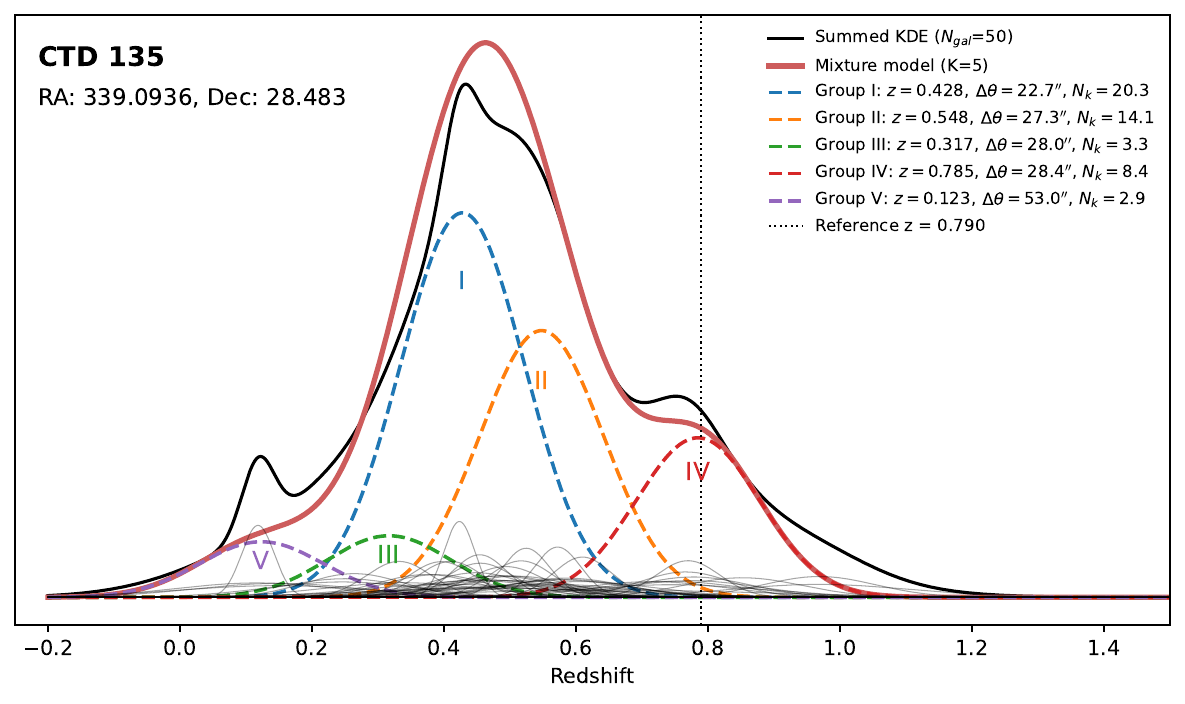}
  \caption{Same as Fig.~\ref{fig:control_spec1}, for the control blazars 1ES~0806$+$524, PKS~0823$+$033, 1ES~1011$+$496, ON~325, W~Comae, RGB~J1415$+$485, OX~183, and CTD~135.} \label{fig:control_spec2}
\end{figure*}

\section{Control MOS blazar plots} \label{sec:mos_plots}

For the MOS control sample (Section~\ref{sec:mos_control}), we show the corresponding SDSS diagnostic plots used in our photometric-redshift analysis. These plots allow a direct comparison between (i) galaxy groups identified via MOS in the literature and (ii) the redshift structures recovered by our SDSS-based GMM procedure.

\begin{figure*}
 \centering
 \includegraphics[width=0.95\linewidth]{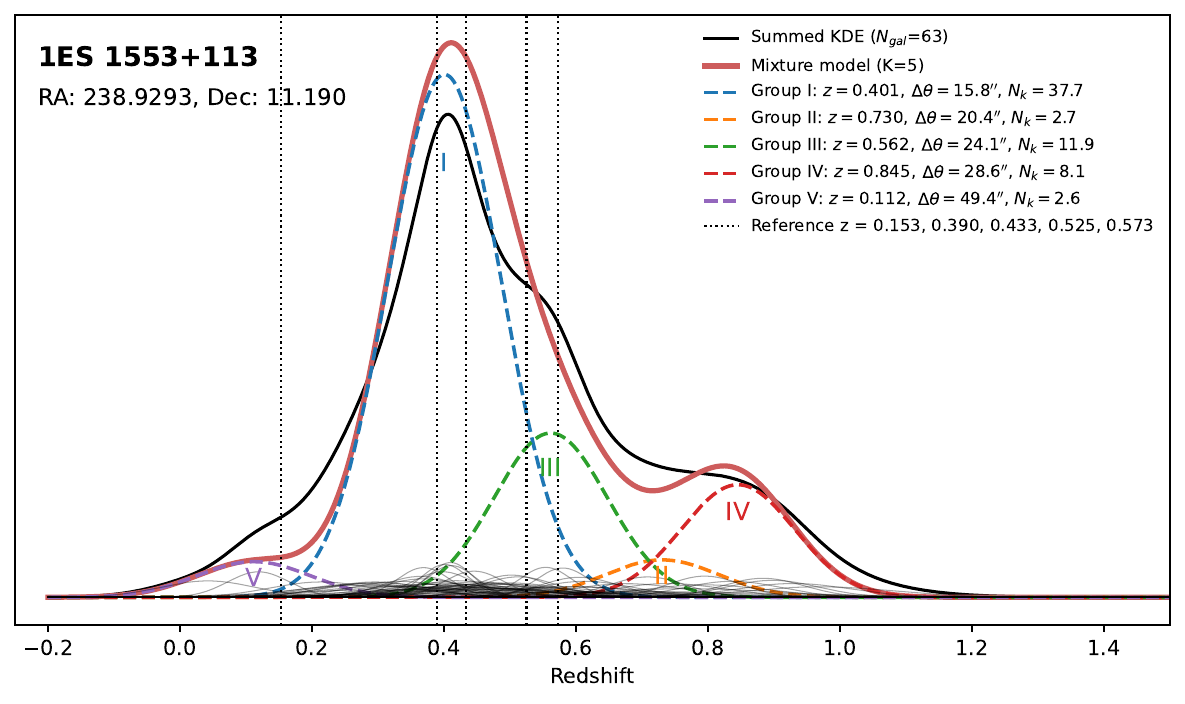}
 \caption{Diagnostic plot (Section~\ref{sec:method}) showing the redshift distribution of SDSS galaxies within $2\arcmin$ of 1ES~1553$+$113. The plot elements are described in Fig.~\ref{fig:pks1424}. 
 }
 \label{fig:1es1553}
\end{figure*} 

\begin{figure*}
 \centering
 \includegraphics[width=0.95\linewidth]{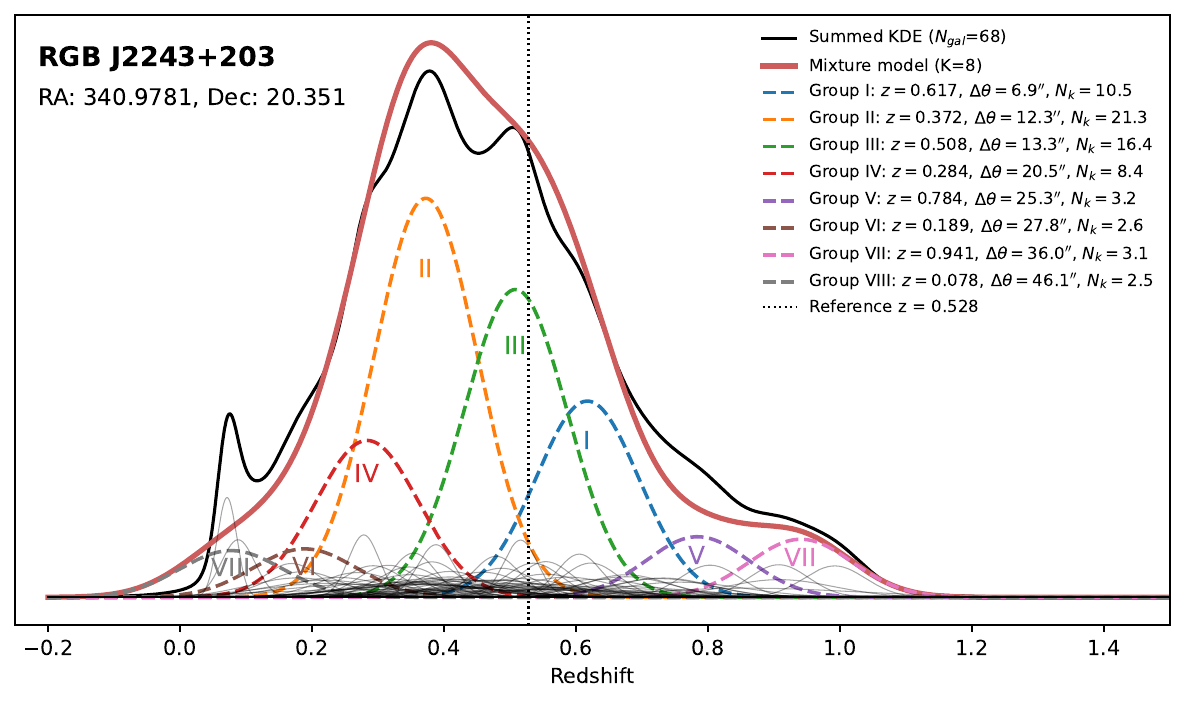}
 \caption{Same as Fig.~\ref{fig:1es1553}, for RGB~J2243$+$203. 
 }
 \label{fig:rgs2243}
\end{figure*} 

\section{PAUS/J-PAS results} \label{sec:paus_plots}

Here we summarise our exploratory comparison between SDSS-based group associations and narrow-band photometric-redshift surveys (Section~\ref{sec:survey_control}). For each blazar within the PAUS or J-PAS footprints, we report (i) the known blazar redshift, (ii) the redshift of the most plausible associated overdensity inferred from PAUS (or J-PAS) narrow-band photometry, and (iii) the corresponding SDSS group-association result obtained using our GMM procedure.

In Table~\ref{tab:pau_redshift}, $N$ denotes the number of galaxies assigned to the associated PAUS/J-PAS overdensity (i.e. contributing to the peak used as the group estimate), $\Delta\theta$ is the angular separation between the blazar and the weighted group centroid, and the SDSS `group order' is the rank of the SDSS GMM centroid by angular separation (I = closest to the blazar).

\begin{table*}
    \caption{Comparison between narrow-band (PAUS/J-PAS) and SDSS group-association results for blazars with known redshifts in the survey footprints. Columns: (1) field (blazar) name; (2) blazar redshift $z$; (3) PAUS/J-PAS group redshift; (4) number of galaxies associated with the PAUS/J-PAS overdensity ($N$); (5) angular separation (in arcseconds) between the blazar and the PAUS/J-PAS weighted group centroid; (6) SDSS GMM group redshift adopted in our analysis; (7) angular separation between the blazar and the SDSS composite centroid of that component; (8) the effective number of galaxies in the group; (9) rank order of the SDSS centroid by angular separation (I = closest).} \label{tab:pau_redshift}
    \centering
    \begin{tabular}{l|cccccccc}   
    \hline
    & & \multicolumn{3}{c}{PAUS} & \multicolumn{3}{c}{SDSS} \\
    Field & $z$ & Group-z & N & $\Delta\theta$ & Group-z & $\Delta\theta$ & Group order \\
    \hline
    BZB~J1401$+$5209 & 0.4819 & 0.480 & 5 & 10 & 0.45 & 18 & 29.1 & I   \\
    BZB~J1406$+$5308 & 0.4577 & 0.458 & 7 & 13 & 0.47 & 26 & 23.2 & II   \\
    SBS~1410$+$530 & 0.0765 & 0.076 & 36 & 63 & 0.12 & 35  & 9.0 & VI   \\
    SBS~1411$+$533 & 0.4562 & 0.454 & 10 & 12 & 0.36 & 7   & 27.2 & I   \\
    PG~1418$+$546 & 0.1525 & 0.152 & 16 & 55 & 0.14 & 21   & 5.1 & VI   \\
    BZB~J1427$+$5409 & 0.1060 & 0.106 & 15 & 32 & 0.15 & 3 & 11.0 & I   \\
    BZB~J1631$+$4217 & 0.4666 & 0.469 & 3 & 14 & 0.44 & 6  & 19.4 & I   \\
    \hline
    \end{tabular}
\end{table*}    

\begin{figure*}
  \centering
  \includegraphics[width=0.49\textwidth]{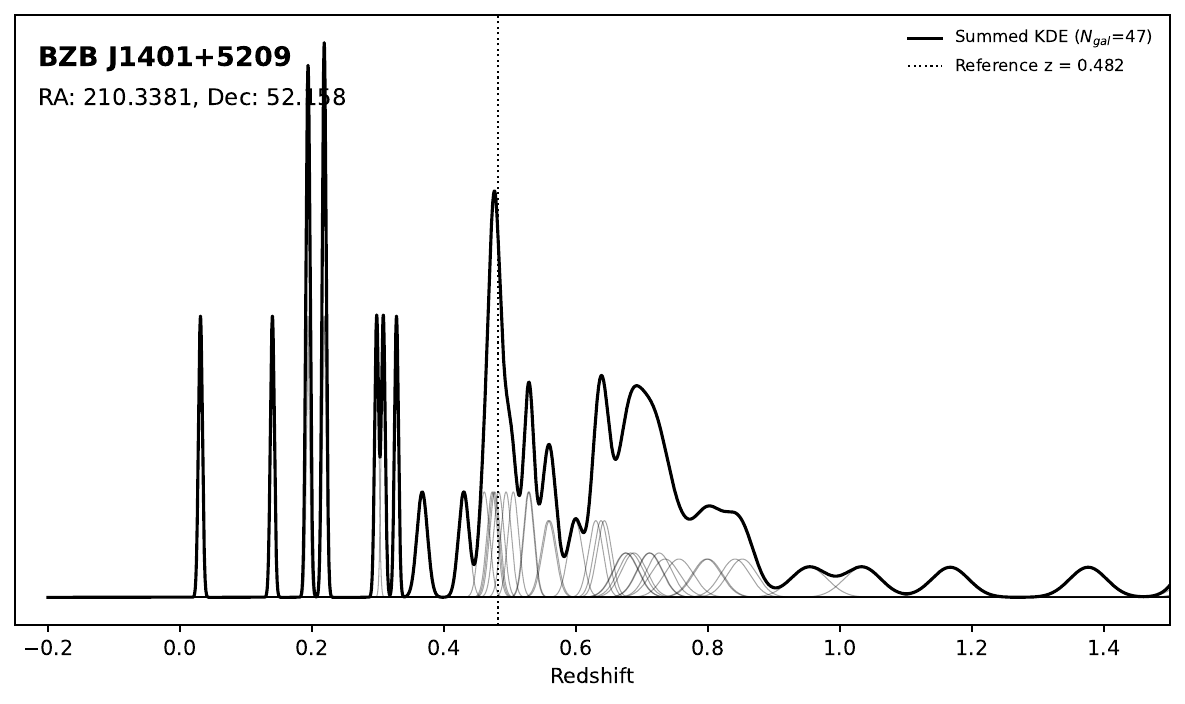}
  \includegraphics[width=0.49\textwidth]{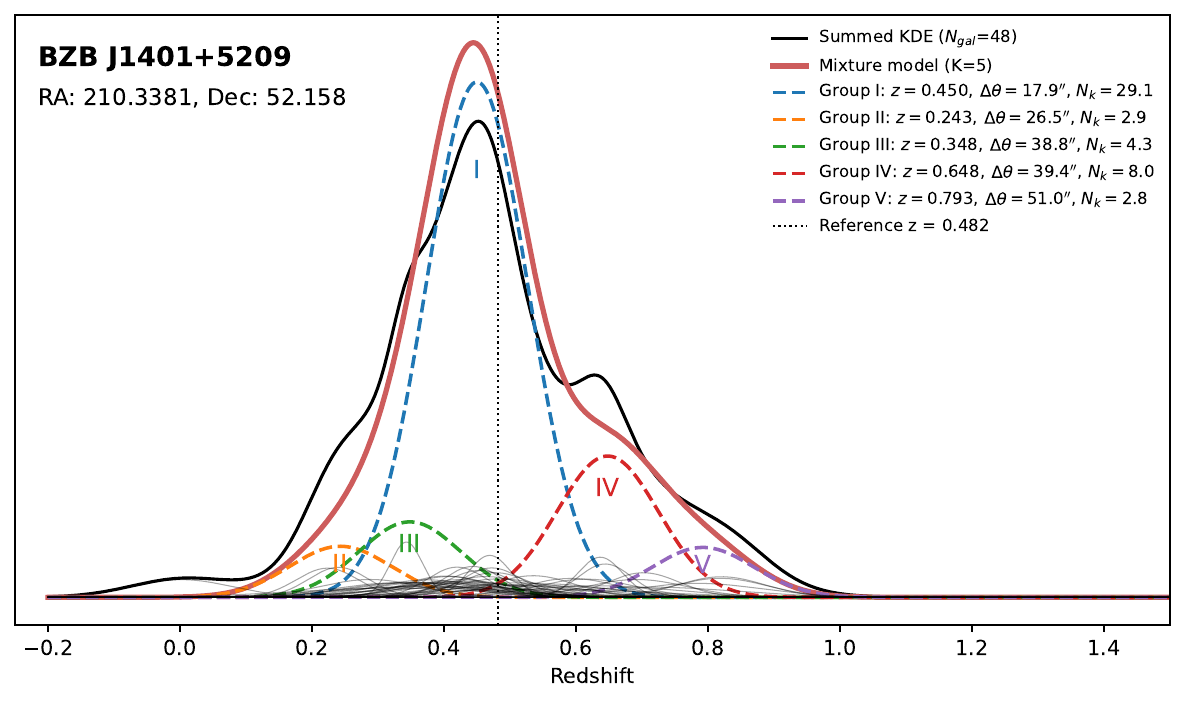}
  \includegraphics[width=0.49\textwidth]{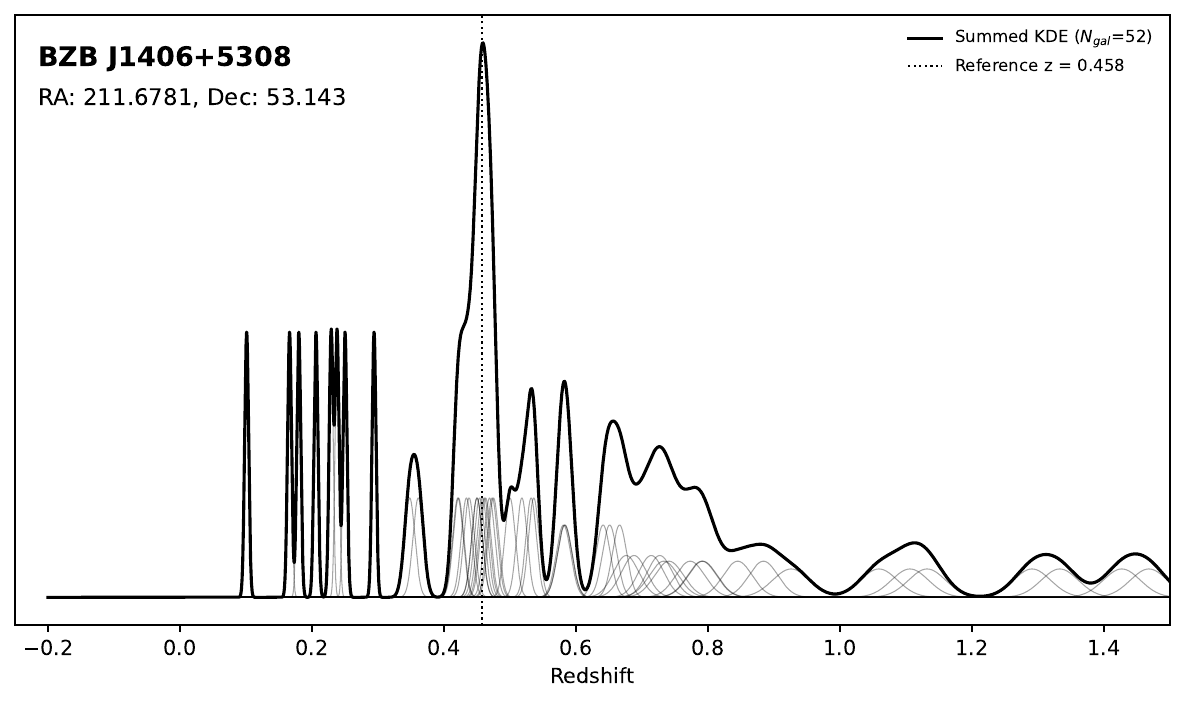}
  \includegraphics[width=0.49\textwidth]{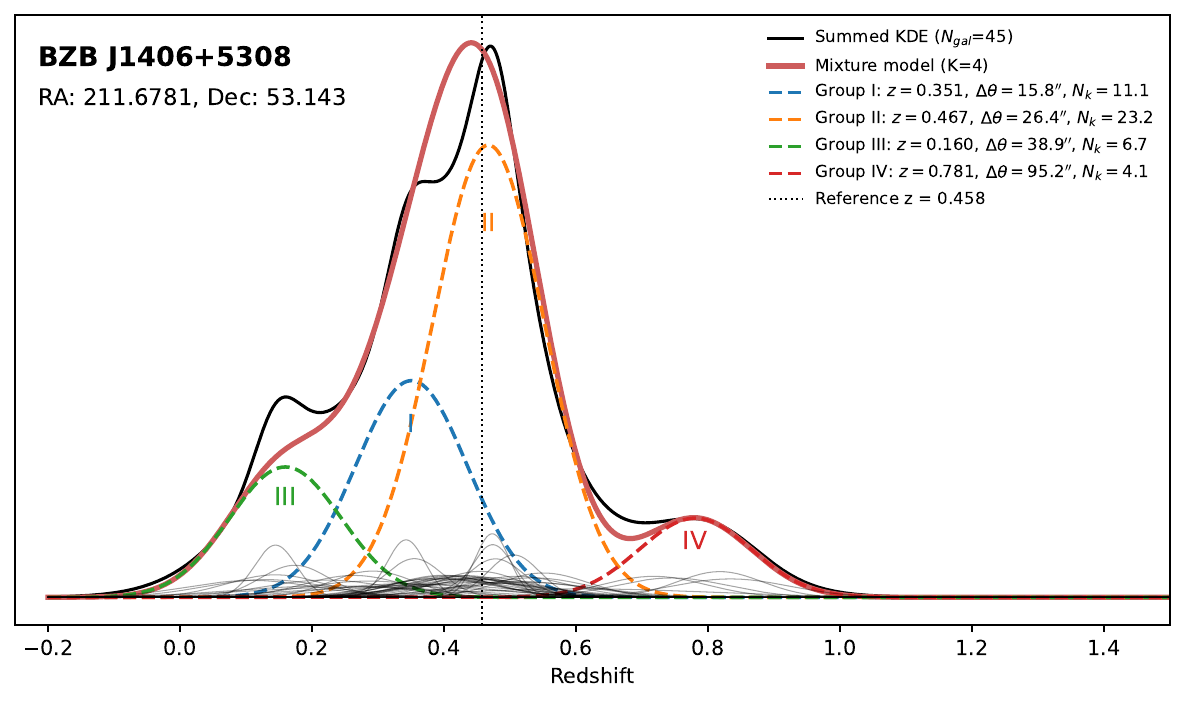}  
  \includegraphics[width=0.49\textwidth]{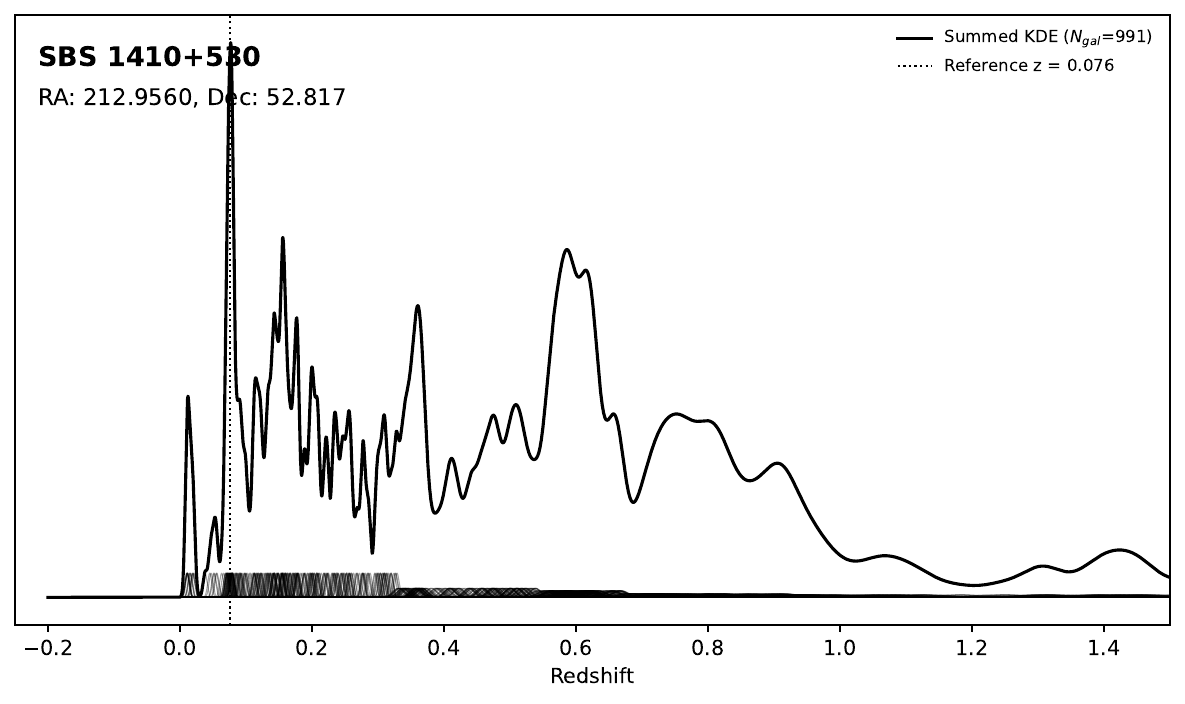}
  \includegraphics[width=0.49\textwidth]{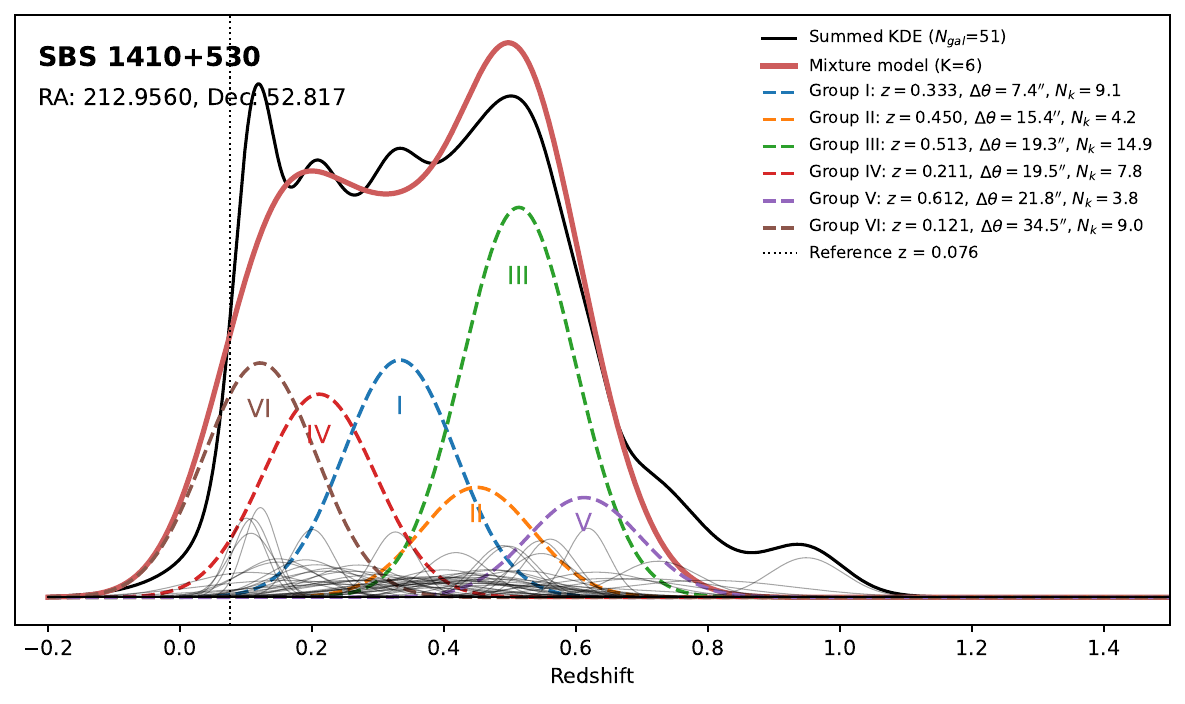}  
  \includegraphics[width=0.49\textwidth]{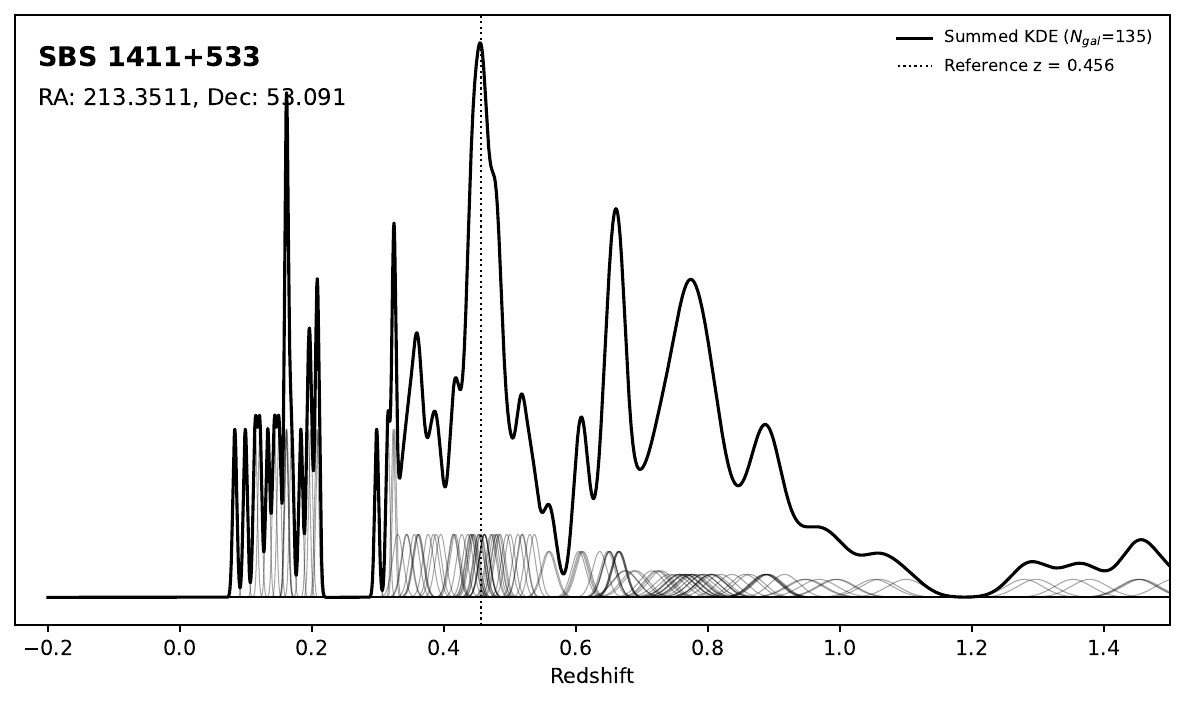}
  \includegraphics[width=0.49\textwidth]{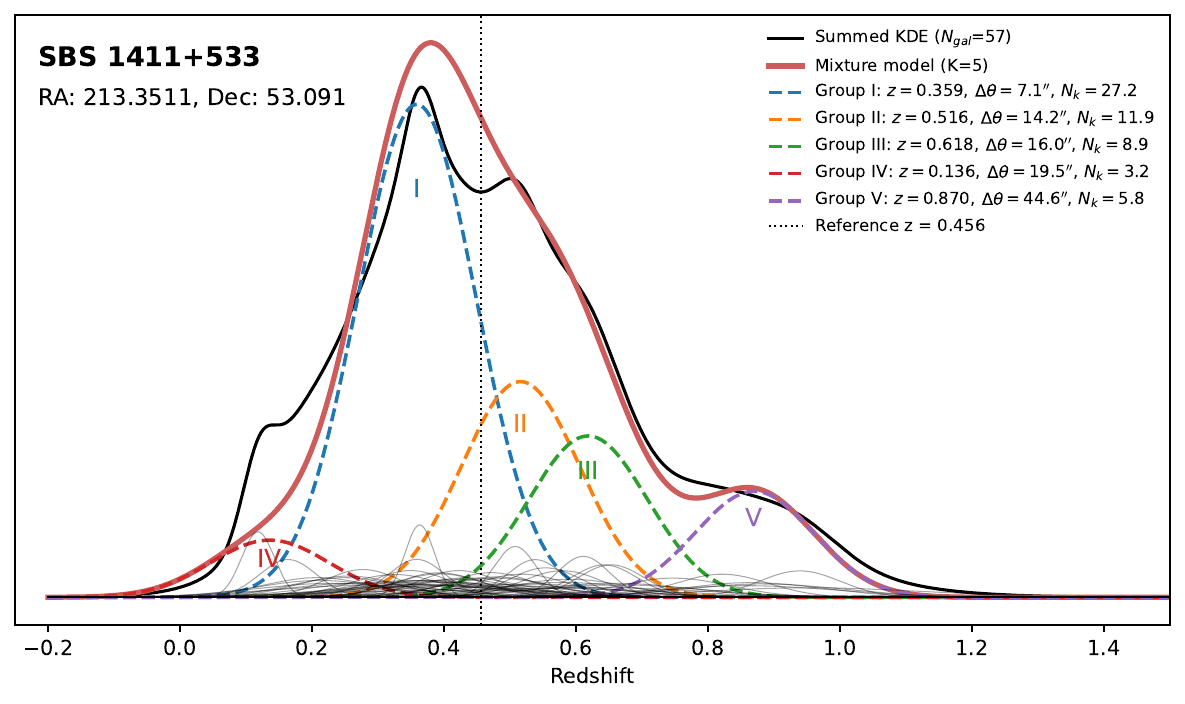}   
  \caption{Comparison of narrow-band and SDSS redshift-structure reconstructions in four fields within the PAUS footprint. Left panels: summed narrow-band redshift distributions (KDEs) for galaxies within a projected 1\,Mpc radius around each blazar (PAUS-MC), with the vertical line indicating the blazar spectroscopic redshift. Right panels: SDSS diagnostic plots (Section~\ref{sec:method}) for the same fields, showing the summed photometric-redshift distribution and the best-fit GMM components.}   \label{fig:paus_results_1}
\end{figure*}

\begin{figure*}
  \centering
  \includegraphics[width=0.49\textwidth]{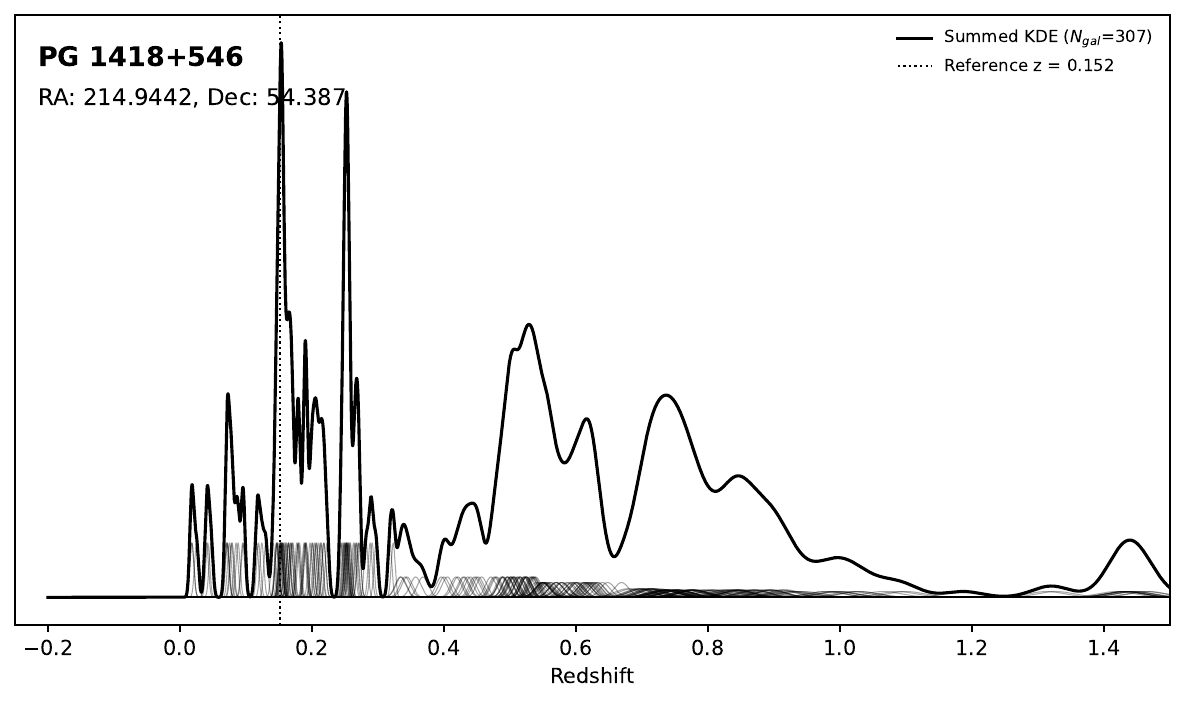}
  \includegraphics[width=0.49\textwidth]{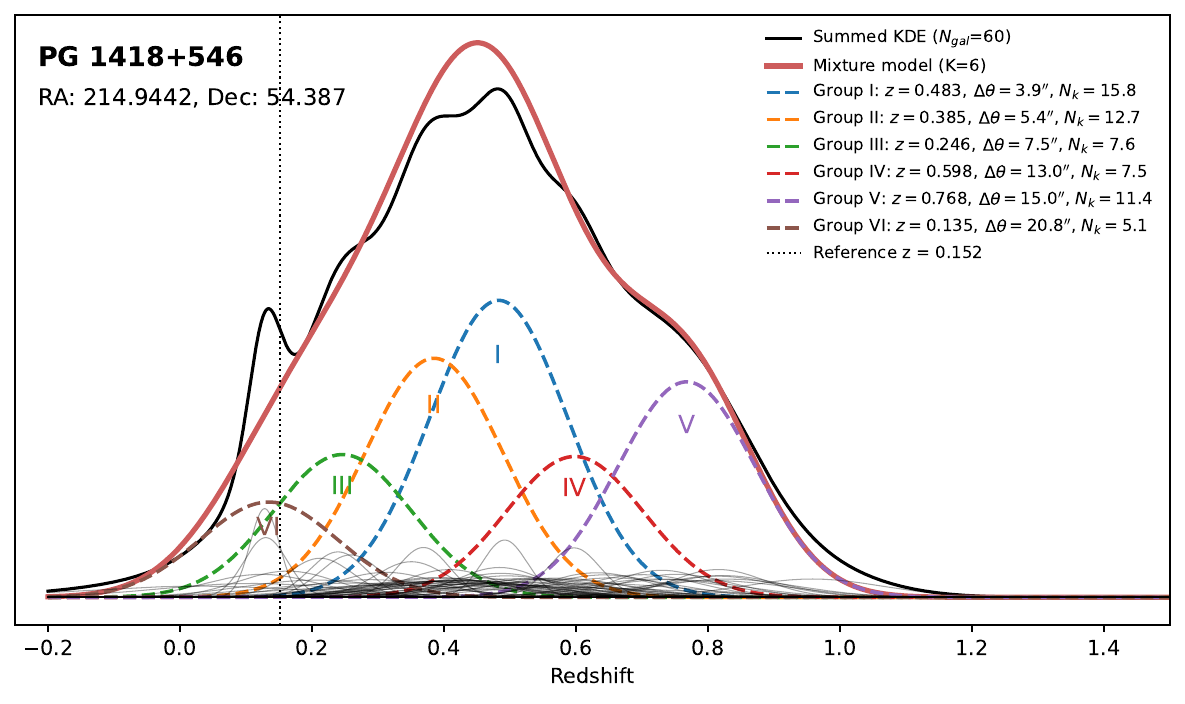}      
  \includegraphics[width=0.49\textwidth]{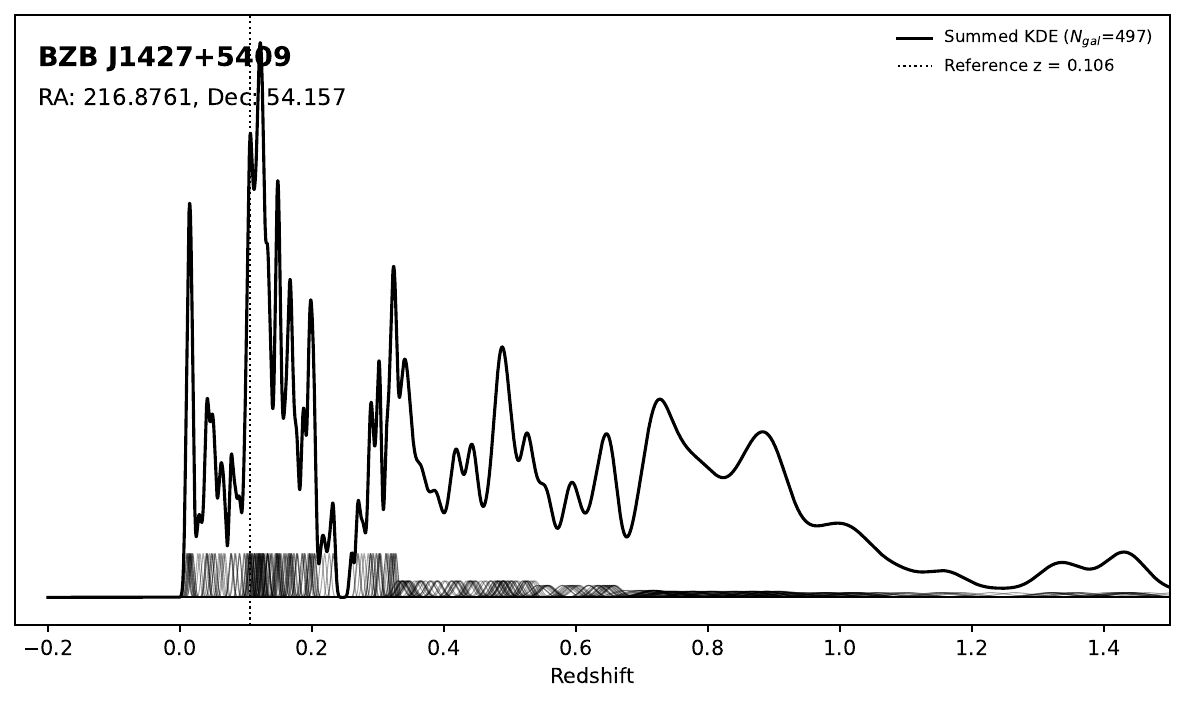}
  \includegraphics[width=0.49\textwidth]{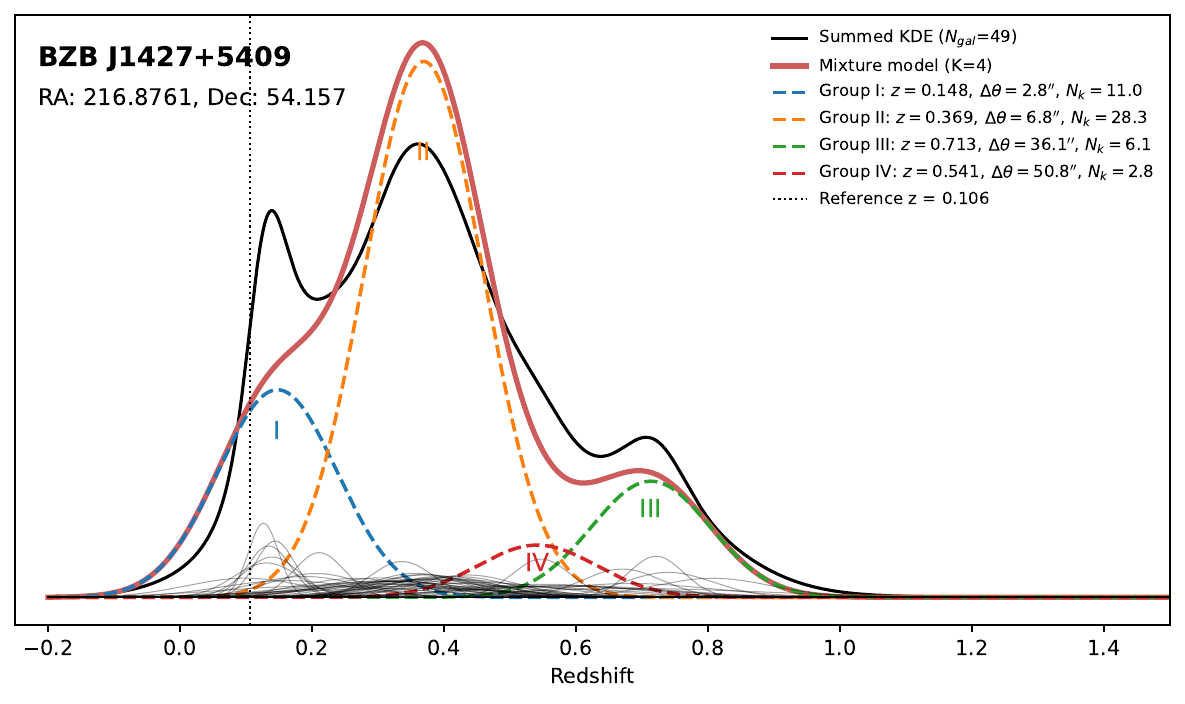}    
  \includegraphics[width=0.49\textwidth]{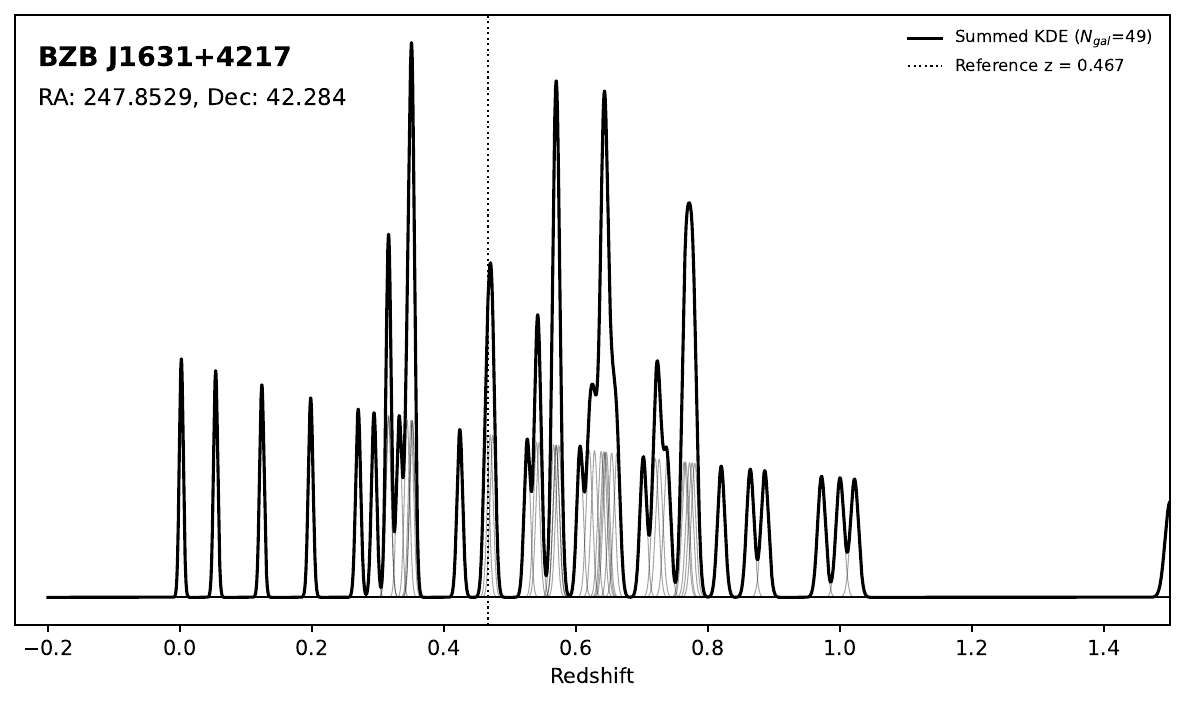}
  \includegraphics[width=0.49\textwidth]{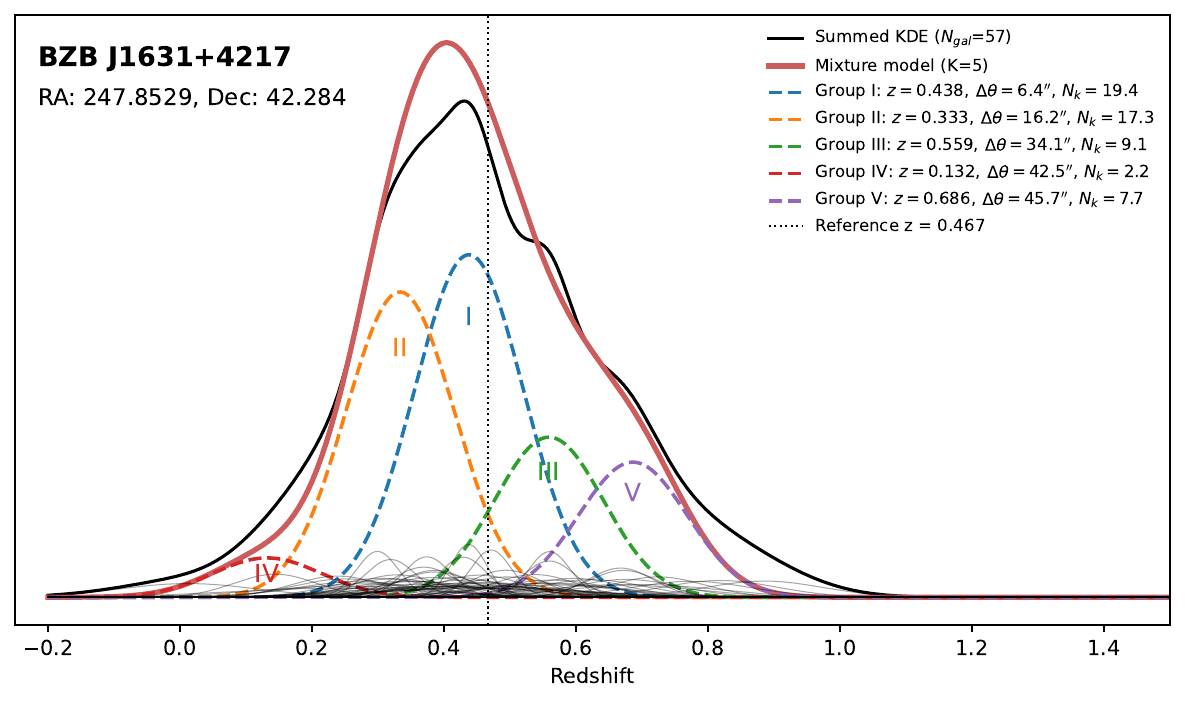}      
  \caption{Same as Fig.~\ref{fig:paus_results_1}, for two additional PAUS fields (PG~1418$+$546 and BZB~J1427$+$5409) and one J-PAS EDR field (BZB~J1631$+$4217). Left panels show narrow-band summed redshift distributions; right panels show the corresponding SDSS diagnostic plots.}   \label{fig:paus_results_2}
\end{figure*}


\bsp	
\label{lastpage}
\end{document}